\documentclass[preprint]{ptephy_v1}

\preprintnumber{arXiv:2306.12437} 
\usepackage{hyperref}

\usepackage{lineno}

\begin{document}

\title{Evaluation of microseismic motion at the KAGRA site based on ocean wave data}


\author[1]{S.Hoshino}
\affil{Faculty of Engineering, Niigata University, 8050 Ikarashi-2-no-cho, Nishi-ku, Niigata City, Niigata 950-2181, Japan \email{f22c036a@mail.cc.niigata-u.ac.jp}}

\author[1]{Y.Fujikawa}
\affil{Faculty of Engineering, Niigata University, 8050 Ikarashi-2-no-cho, Nishi-ku, Niigata City, Niigata 950-2181, Japan}

\author[1]{M.Ohkawa}
\affil{Faculty of Engineering, Niigata University, 8050 Ikarashi-2-no-cho, Nishi-ku, Niigata City, Niigata 950-2181, Japan}

\author[2]{T.Washimi}
\affil{Gravitational Wave Science Project (GWSP), Kamioka branch, National Astronomical Observatory of Japan (NAOJ), Kamioka-cho, Hida City, Gifu 506-1205, Japan\email{tatsuki.washimi@nao.ac.jp}}

\author[3]{T.Yokozawa}
\affil{Institute for Cosmic Ray Research (ICRR), KAGRA Observatory, The University of Tokyo, Kamioka-cho, Hida City, Gifu 506-1205, Japan}


\begin{abstract}%
The microseismic motion, ambient ground vibration caused by ocean waves, affects ground-based gravitational wave detectors. 
In this study, characteristics of the ocean waves including seasonal variations and correlation coefficients were investigated for the significant wave heights at 13 coasts in Japan. The relationship between the ocean waves and the microseismic motion at the KAGRA site was also evaluated.
As a result, it almost succeeded in explaining the microseismic motion at the KAGRA site by the principal components of the ocean wave data. 
One possible application of this study is microseismic forecasting, an example of which is also presented
\end{abstract}

\subjectindex{xxxx, xxx}

\maketitle

\section{Introduction}\label{sec:Introduction}
Gravitational waves are ripples in space-time propagating at the speed of light and their direct observation is a key probe in advanced astronomy. 
The first successful detection was performed in 2015 by the advanced Laser Interferometer Gravitational-Wave Observatory (LIGO, USA)~\cite{LIGO paper, GW150914}, and the first simultaneous detections by LIGO and Virgo (Italy) were performed in 2017~\cite{Virgo paper, GW170814, GW170817}.
KAGRA is a laser interferometric gravitational wave detector with 3~km arms in Japan~\cite{2020_Overview_of_KAGRA}. 
Two solo observation runs were conducted in 2015 and 2018~\cite{iKAGRA, bKAGRA phase1}, and the first international joint observation run (O3GK) with GEO 600 in Germany was conducted from April 7–21, 2020~\cite{PTEP013F01, EDSU2020}. 
KAGRA has two unique features compared with other kilometer-scale detectors in the world: 
(1) it is constructed underground at Kamioka to reduce the ground vibration noise,
(2) the test mass mirrors are cooled down to reduce thermal noise.
To attain and maintain the working point of the detector, all mirror positions, angles, and motions must be controlled.
When external disturbances occur, maintaining these controls becomes difficult, and the resonant state is broken. Consequently, the gravitational wave observations must be stopped. This state is called "lock loss" and a reduction in the lock loss rate is important for performing meaningful observations.
During the O3GK period, it was occasionally difficult to keep the KAGRA interferometer in a locked state when microseismic motions, which are ground vibrations with a frequency range of about 0.1–0.5 ~Hz induced by ocean waves, were significant~\cite{EDSU2020, galaxies1601430, Fujikawa}.
The mechanism of microseismic motion excited by ocean waves was derived by Longuet-Higgins~\cite{Longuet-Higgins} and approximated using a non-linear equation extended by Hasselmann~\cite{Hasselmann}. 
This approximation was evaluated using the normal-mode equation derived by Tanimoto to obtain negligible errors for ground vibrations occurring in the ocean to a depth of approximately 1~km~\cite{Tanimoto, microseisms}. 
Microseismic motion is related to the amplitude and period of the waves, with the period of the ground motions being approximately half the period of the waves, and the magnitude of the motions being derived from the energy of the waves, as shown by Bromirski {\it et al.}~\cite{Ocean wave height}.

In this study, we investigate the relationship between the microseismic motion at the KAGRA site and the activity of the ocean waves around Japan; Although this relationship has been expected qualitatively, it has not been studied quantitatively. Sec.~\ref{sec:seis} explains the details of the seismometers installed at the KAGRA site and the behavior of the measured microseismic motion. In Sec.~\ref{sec:wave}, the characteristics of the ocean waves around Japan are discussed based on public data. In Sec.~\ref{sec:main}, we propose an approximate equation connecting the microseismic motion at the KAGRA site to the wave height at various coasts in Japan. Finally, in Sec.~\ref{sec: prospects}, the results of this study are summarized and its useful application, a forecast of future microseismic motion, is shown.

\section{Characterizations for the microseismic motion at the KAGRA site}\label{sec:seis}
\begin{figure}[htbp] \centering
    \includegraphics[clip, width=7.5cm]{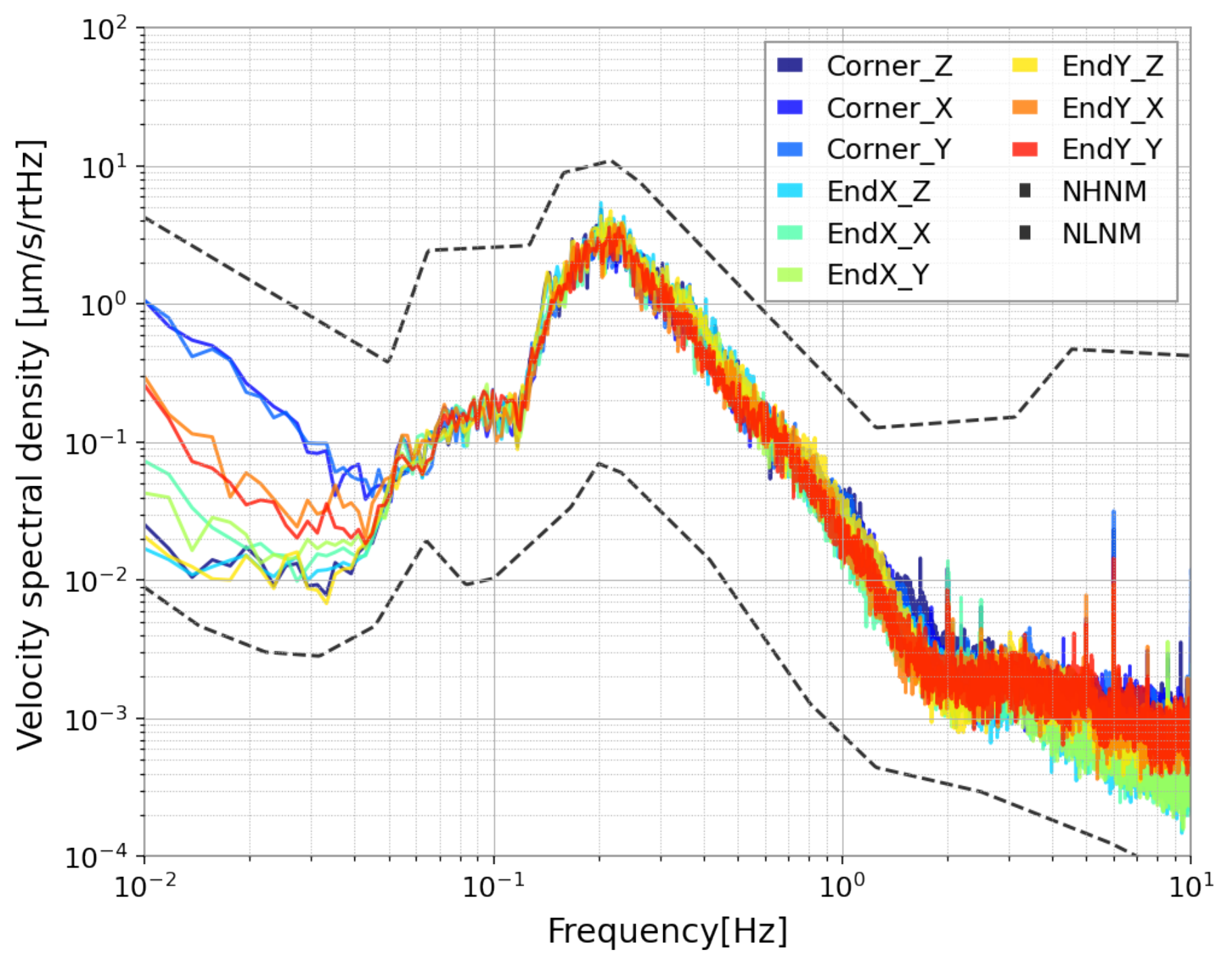}
    \includegraphics[clip, width=7.5cm]{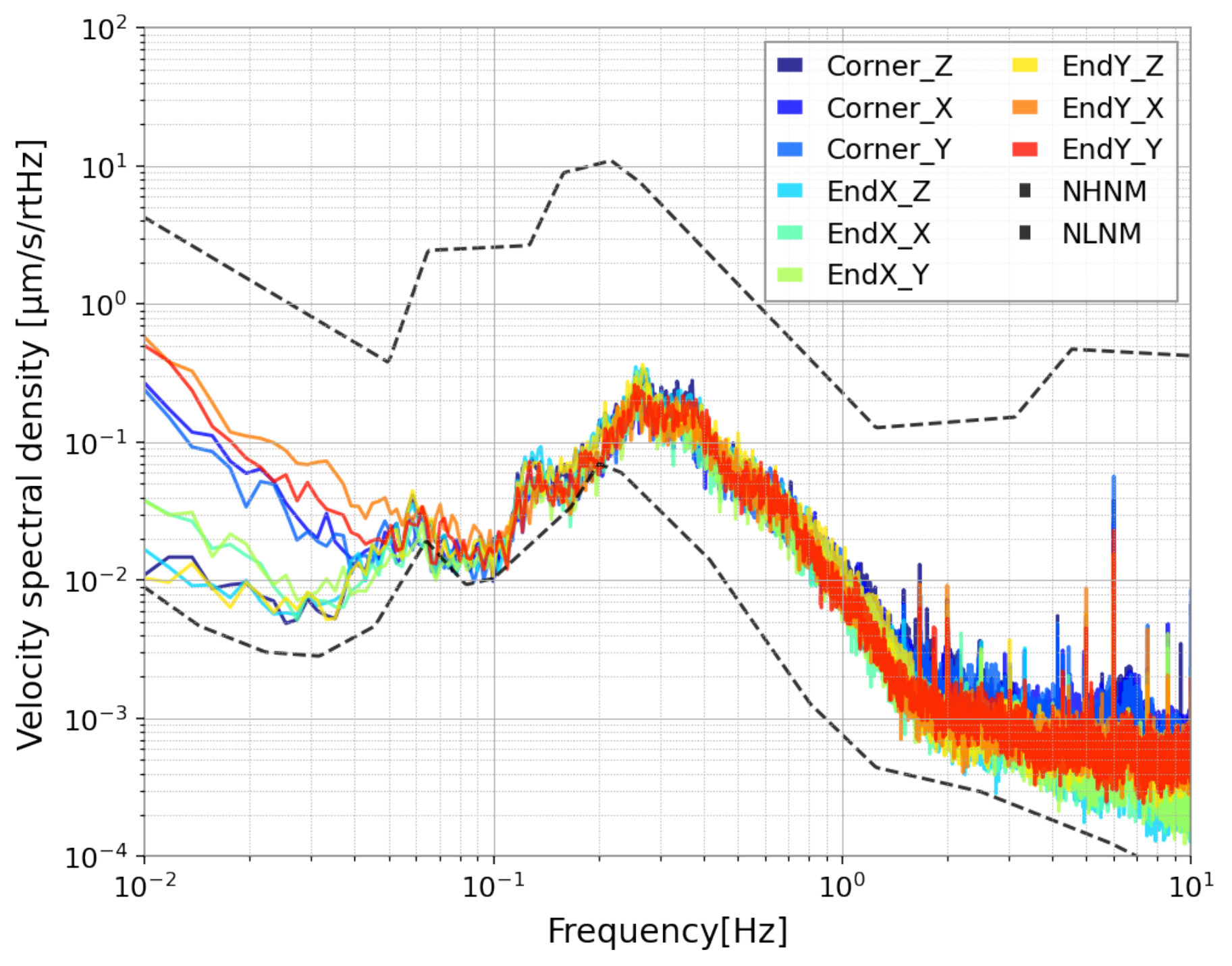}
    \caption{Amplitude spectral densities (ASDs) of the ground velocity, for each location (Corner, X-end, Y-end) and each direction (X, Y, Z) at the KAGRA site. The measurement time is 4096 seconds on February 18 (left) and June 10 (right) in 2020. 
    Black dashed lines represent Peterson's high/low seismic noise model~\cite{Noise model}.}
    \label{fig:SeisASD}
\end{figure}

In this section, we explain the information about the seismometers used in this study and the characteristics of their data. 
To monitor the environments around the KAGRA interferometer, several sensors were installed at the experimental site and continuously logged using the KAGRA DAQ system~\cite{KAGRA underground environment}.
Three seismometers were placed at each end and corner station, and  the horizontal axes were aligned with the orientations of the arms~\cite{KAGRA experimental site}.
The seismometers used are the Trillium 120QA from \textit{Nanometrics Inc.}, which are sensitive to ground velocity in three directions from 0.01~Hz to 10~Hz. 
Fig.~\ref{fig:SeisASD} shows examples of the amplitude spectral density (ASD) of the ground velocity of these seismometers in each location and direction. 
In this figure, two different days are shown to compare high- and low-noise days, with black dashed lines representing Peterson’s high/low seismic noise model~\cite{Noise model}. 
The significant peak at 0.1--0.5~Hz corresponding to the microseismic motion is seen in all ASDs and its amplitude and structure are almost the same in the stations and the directions for both days. 
The amplitude below 0.05~Hz varies in the channels, and this behavior is assumed to be due to atmospheric pressure ~\cite{tonga, Alessandro}. 
Based on these results, for simplicity, the vertical signal of the seismometer located at the corner station was used to represent  ground vibration in this study.

\begin{figure}[htbp] \centering
    \includegraphics[clip,width=16cm]{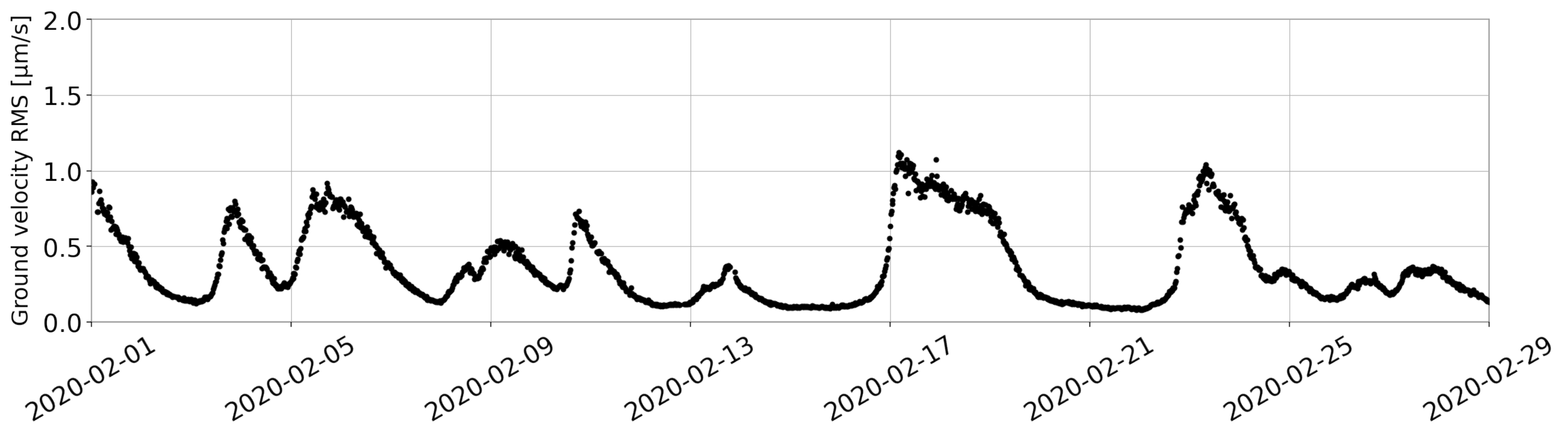}
    \caption{The root mean square of the microseismic motion (0.1--0.5~Hz) at the KAGRA site during  February 2020. All data for 2020 are shown in the \ref{sec:OneYear}.}
    \label{fig:SeisRMS_Feb}
\end{figure}

The band-limited root mean square (BLRMS) of the seismometer signal was used to evaluate the time dependence of microseismic motion at the KAGRA site. 
BLRMS is the root mean square every 20 minutes for the time series data filtered using a bandpass filter (\texttt{TimeSeries.bandpass} in \texttt{gwpy 2.1.4}) from 0.1~Hz to 0.5~Hz, to limit the frequency band. 
Fig.~\ref{fig:SeisRMS_Feb} shows an example from February 2020. All the data for 2020 are shown in \ref{sec:OneYear}.
In the O3GK run, the KAGRA interferometer was difficult to keep locked when the BLRMS value was above $0.3\ \mathrm{\mu m/s}$ and was impossible to keep locked when the BLRMS value was over $0.5\ \mu\mathrm{m/s}$~\cite{Fujikawa}. 

\begin{figure}[htbp] \centering
    \includegraphics[clip,width=16cm]{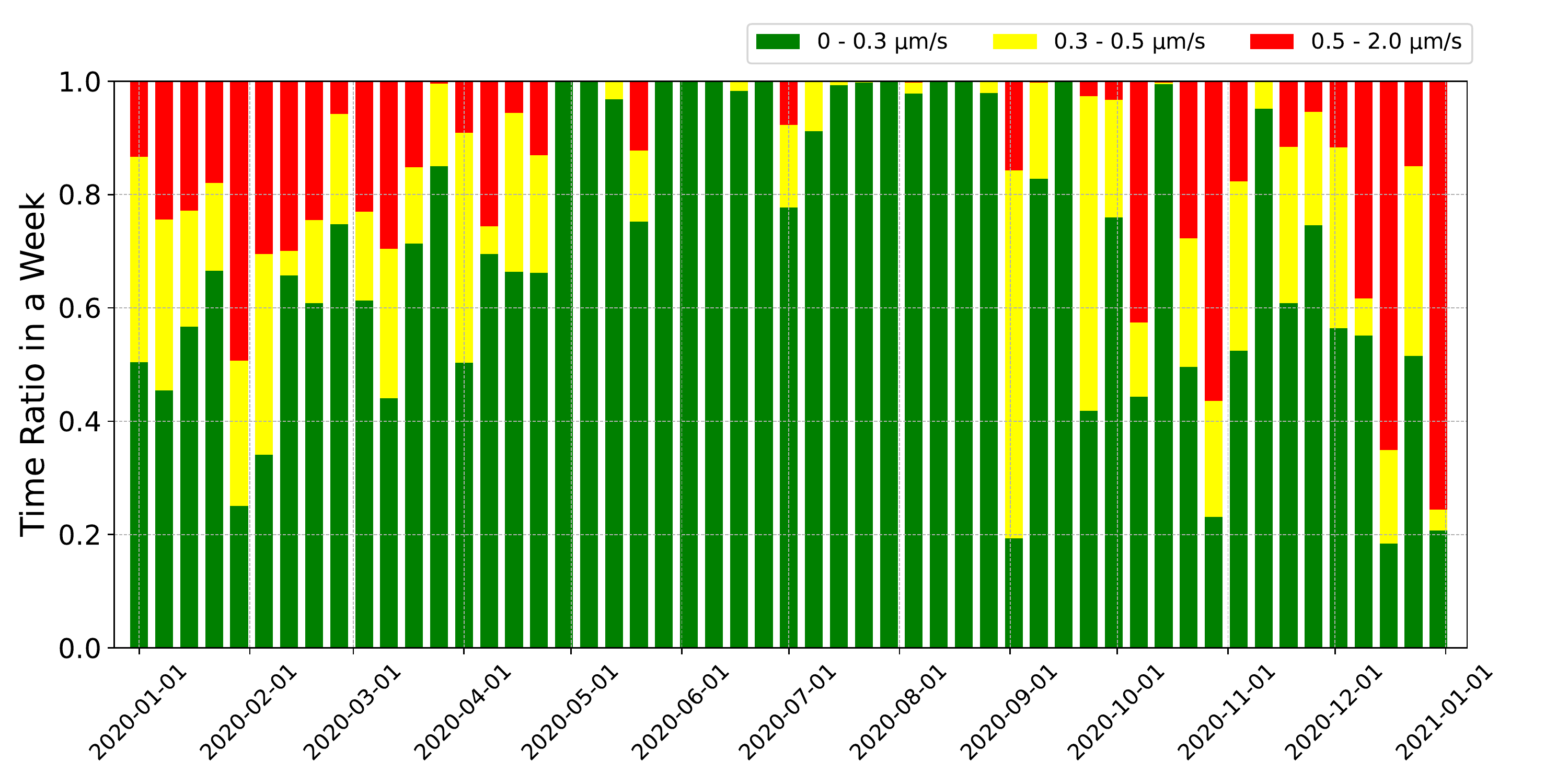}
    \caption{The time ratio for every week in 2020, classified by the microseismic motion level into three ranges:  below $0.3\ \mu\mathrm{m/s}$ (green), 0.3--0.5~$\mu\mathrm{m/s}$ (yellow), and above $0.5\ \mu\mathrm{m/s}$ (red). 
    This classification is based on the lock-state of KAGRA in the O3GK~\cite{Fujikawa}. }
    \label{fig:StackGragh}
\end{figure}

Fig.~\ref{fig:StackGragh} shows the time ratio for every week in 2020, classified by the microseismic motion level into three ranges: below $0.3\ \mu\mathrm{m/s}$ (green), between $0.3\ \mu\mathrm{m/s}$ and $0.5\ \mu\mathrm{m/s}$ (yellow), and above $0.5\ \mu\mathrm{m/s}$ (red). 
The microseismic motion increased from winter to the beginning of spring (December--March) and remained stable at small values in summer. 
It also shows large values at the beginning of autumn (September and October) owing to typhoons.

\section{Characterizations for the ocean waves around Japan}\label{sec:wave}
\begin{figure}[htbp] \centering
    \includegraphics[clip,width=15cm]{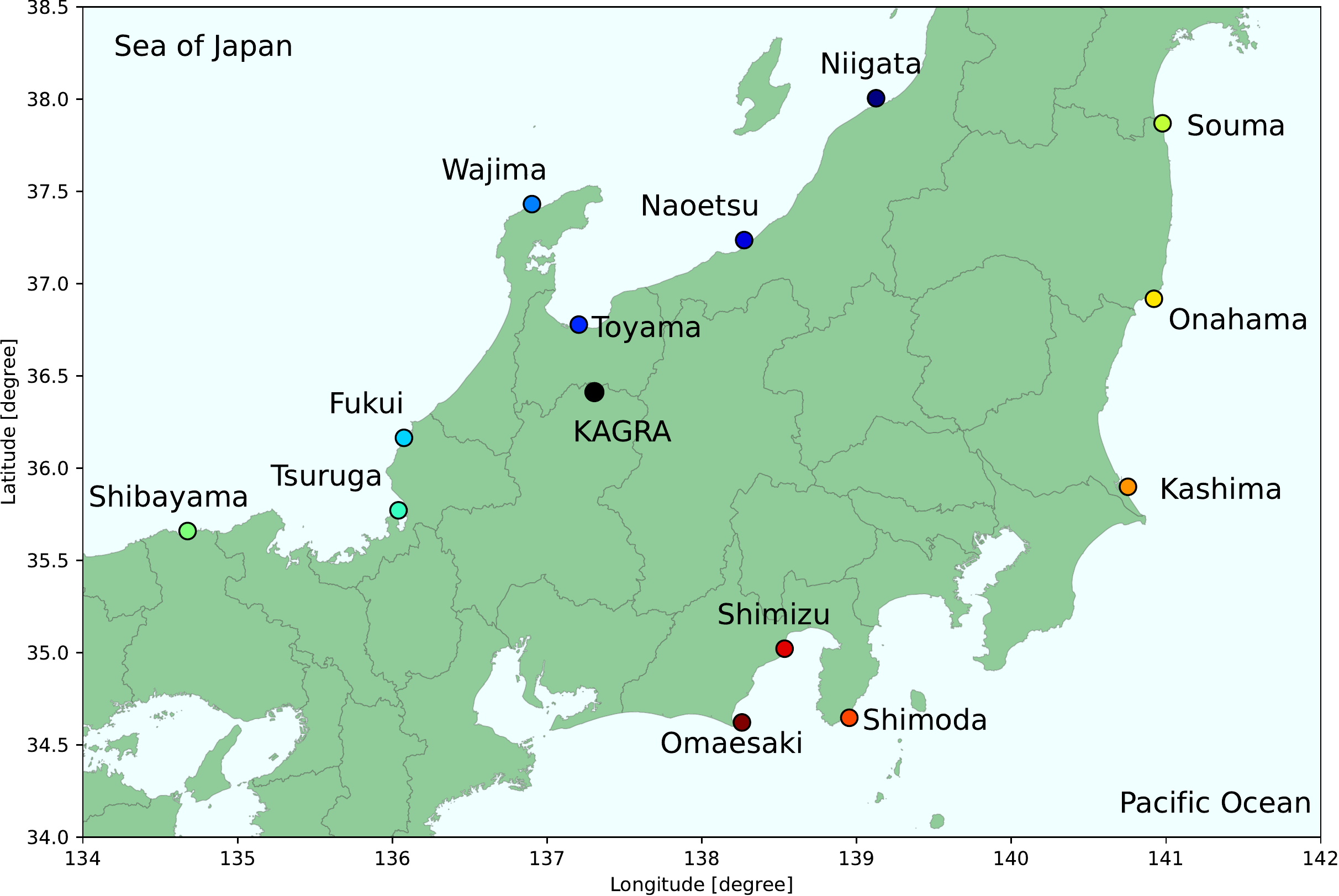}
    \caption{Locations of the KAGRA (black marker) and the NOWPHAS observatories used in this study (color markers): Niigata, Naoetsu, Toyama, Wajima, Fukui, Tsuruga, and Shibayama on the Sea of Japan side, and Soma, Onahama, Kashima, Shimoda, Shimizu, and Omaesaki on the Pacific side.}
    \label{fig:OceanMap}
\end{figure}

Significant wave height (SWH, $H_{1/3}$) is the average of the highest one-third (33\%) of waves (measured from trough to crest) that occur in a given period~\cite{SWH} and is widely used as an indicator of the strength of ocean waves. 
Wave data are provided by {\it the Nationwide Ocean Wave information network for Ports and HArborS} (NOWPHAS) operated by {\it the Port and Harbor Bureau, Ministry of Land, Infrastructure, Transport and Tourism}, Japan~\cite{NOWPHAS}. 
NOWPHAS data are measured every 20~minutes using the zero-up-cross method~\cite{ZeroUpCross}.
Seven sites on the Sea of Japan coast (Niigata, Naoetsu, Toyama, Wajima, Fukui, Tsuruga, and Shibayama) and six sites on the Pacific side (Soma, Onahama, Kashima, Shimoda, Shimizu, and Omaesaki), as shown in Fig.~\ref{fig:OceanMap}, are selected for use in this study from the NOWPHAS data. Among the selected sites, the closest one is Toyama, about 45 km, and the farthest one is Soma, about 350 km away from the KAGRA site.
Fig.~\ref{fig:Sec3TimeSeries} shows the time series of the SWH at these 13 sites in February 2020 and 
Fig. A1 (in Appendix A.) shows them throughout 2020.

\begin{figure}[htbp]\centering
    \includegraphics[clip,width=16cm]{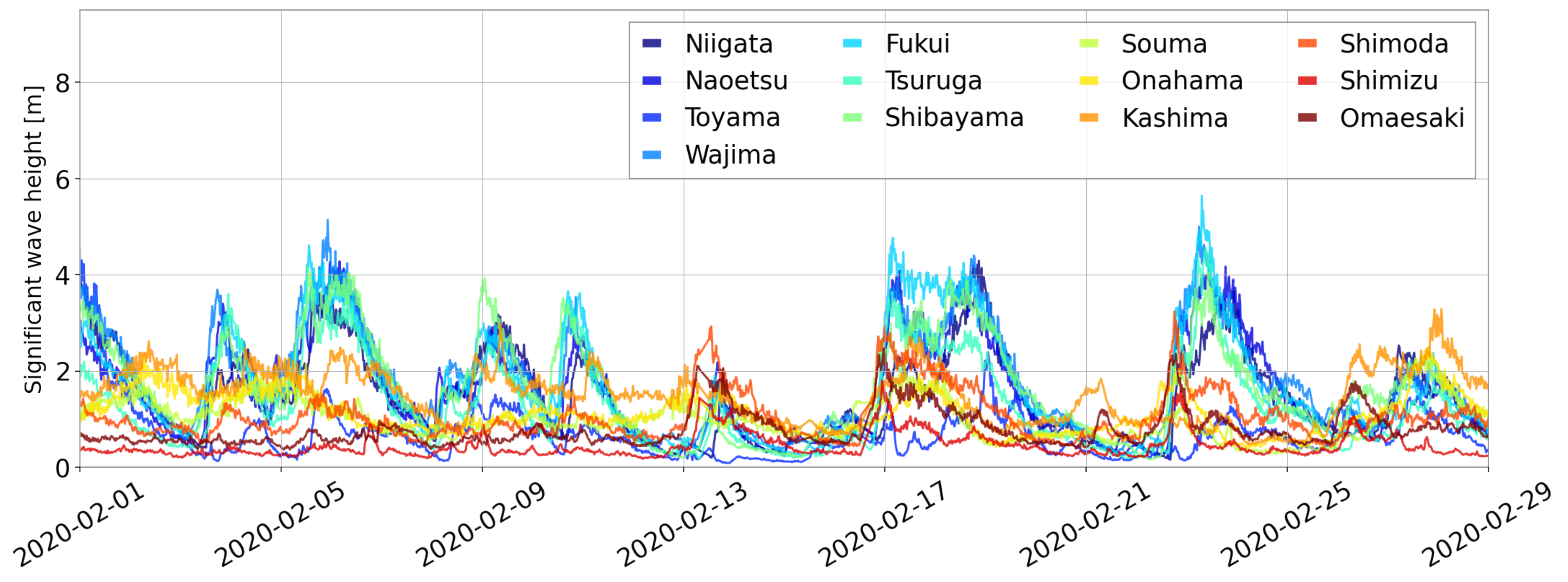}   
    \caption{Time series of the significant wave heights (SWH) during February 2020, provided by NOWPHAS~\cite{NOWPHAS}.
    All data for 2020 are shown in the \ref{sec:OneYear}.}
    \label{fig:Sec3TimeSeries}
\end{figure}

\begin{figure}[htbp]    \centering
    \includegraphics[clip,width=14cm]{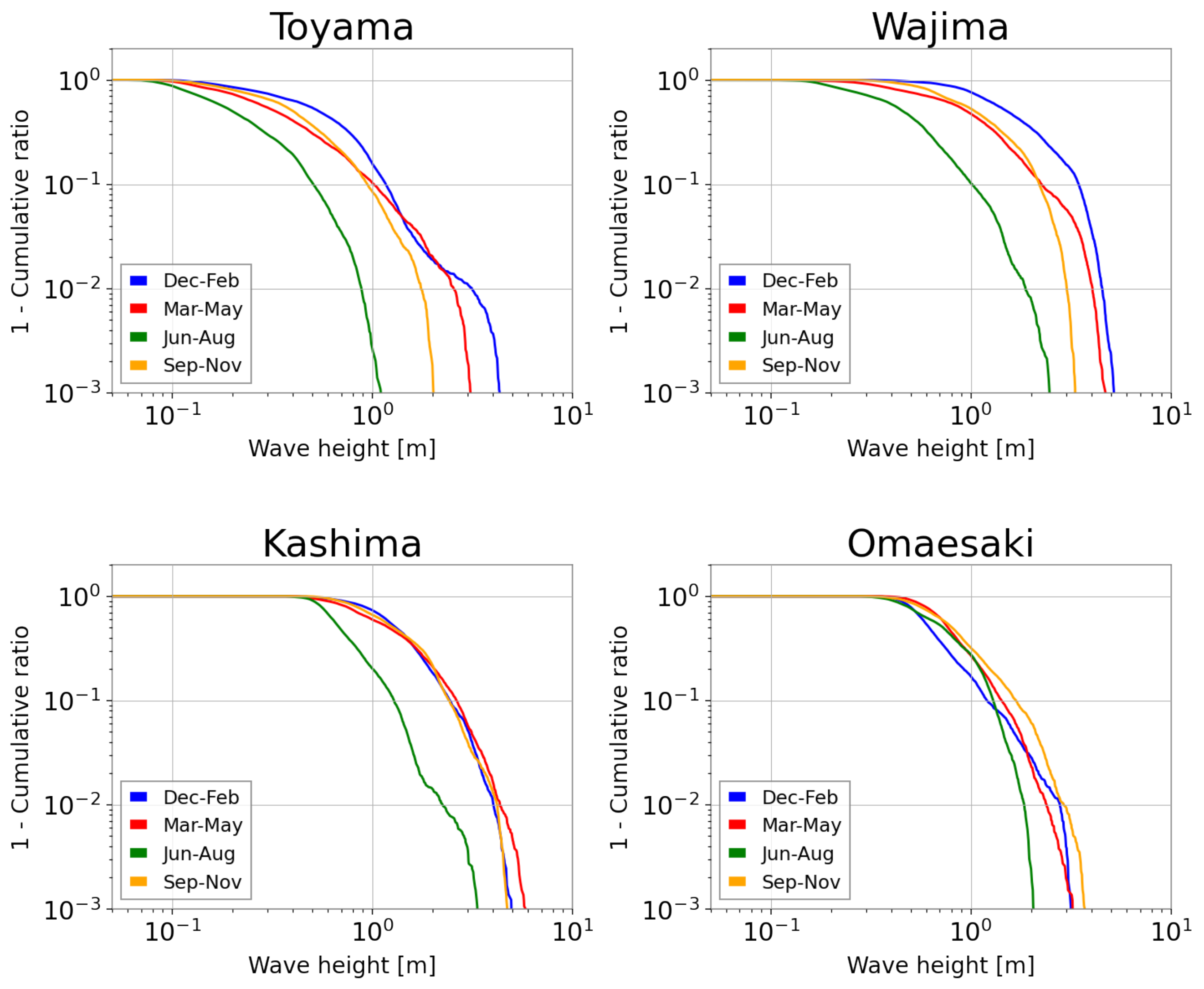}
    \caption{The cumulative distribution functions of the SWH, which are normalized to ratios and subtracted from 1 to describe the behaviors at larger values of the SWH, at the four sites (Toyama, Wajima, Kashima, and Omaesaki) for 2020, for each season: 
    winter (January, February, and December), 
    spring (from March to May), summer (from June to August), and autumn (from September to November).}
    \label{fig:WaveSeason}
\end{figure}

\begin{figure}[htbp]    \centering
    \includegraphics[clip,width=15cm]{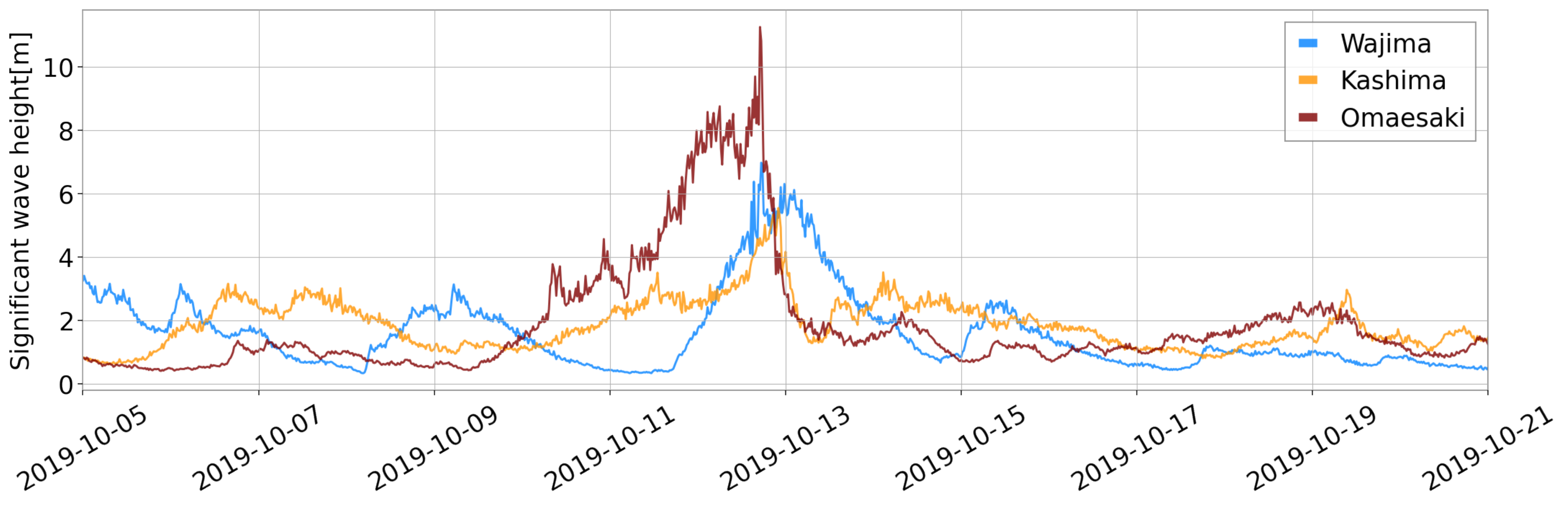}
    \caption{Time series of the significant wave heights (SWH) at three sites (Wajima, Kashima, and Omaesaki), during the typhoon period in October 2019.}
    \label{fig:WaveHeight2019}
\end{figure}

\begin{figure}[htbp]
    \centering
    \includegraphics[clip,width=16cm]{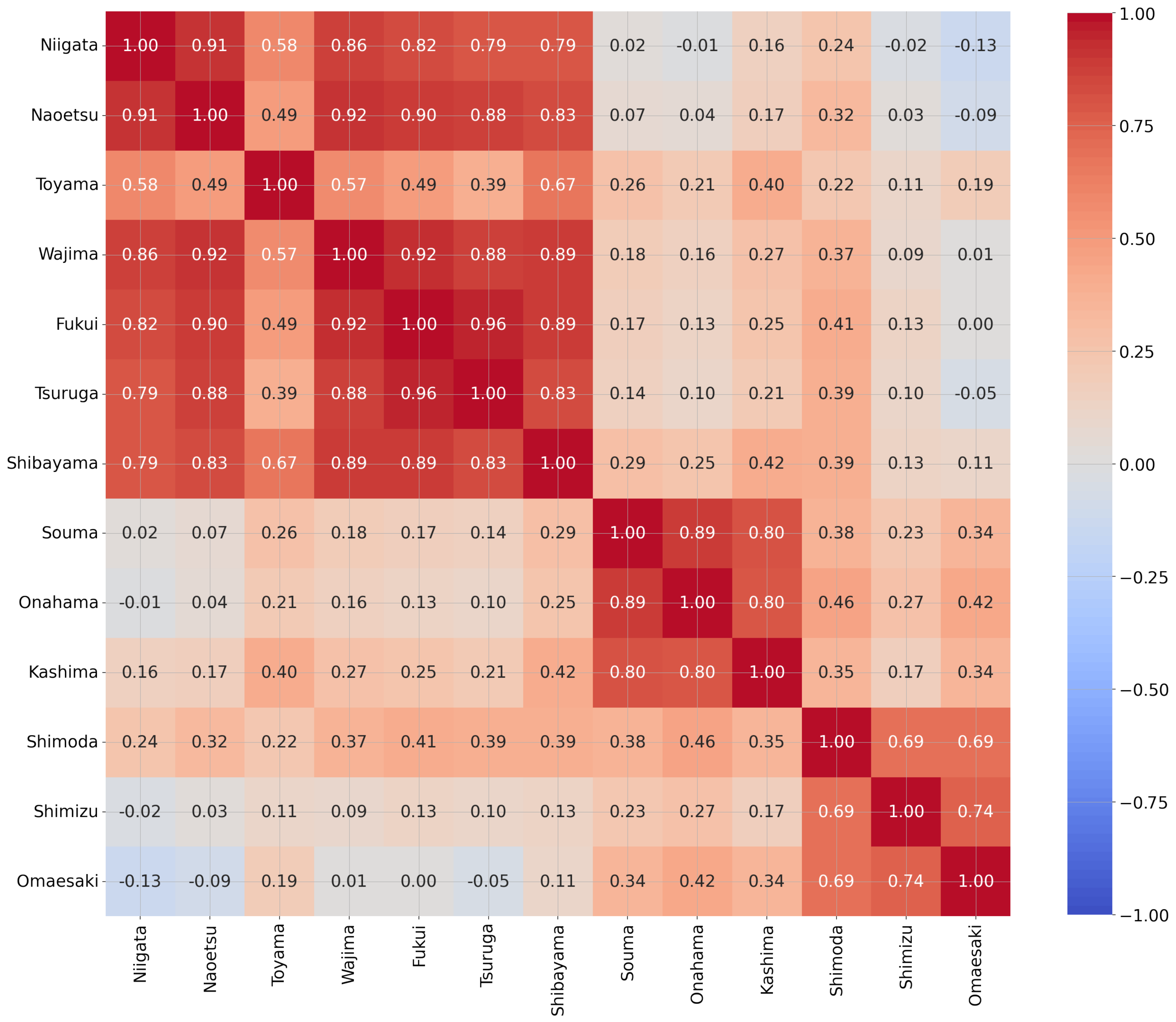}
    \caption{The Pearson correlation coefficient of the significant wave height (SWH) between the 13 coasts of Japan. The SWH data is collected every 20 minutes in 2020 and provided by the NOWPHAS~\cite{NOWPHAS}.}
    \label{fig:OceanCorr}
\end{figure}

Fig.~\ref{fig:WaveSeason} shows the cumulative distribution functions of the SWH, which are normalized to ratios and subtracted from 1 to describe the behaviors at larger values of the SWH, at the four sites (Toyama, Wajima, Kashima, and Omaesaki) calculated every three months.
\footnote{Only four sites are shown in Fig.~\ref{fig:WaveSeason} because the SWH at each site has a strong positive correlation with the other sites and shows similar trends. See Fig.~\ref{fig:OceanCorr} for details.}
The waves are relatively larger in the winter seasons at Wajima and Toyama, which is consistent with the fact that the wind in the Sea of Japan is strongly affected by seasonal wind and becomes stronger in these seasons. 
Seasonal winds blow from the northwest during winter and from the southeast during summer in Japan. 
Toyama Bay is traditionally known as a "quiet bay", and its SWH values are smaller than those of Wajima, even though these sites are close. 
In Kashima, wave activities seem to be at the same level, except during the summer period.
At Omaesaki, there was little change throughout the year approximately 90\% of the probability.
Typhoons typically approach Japan between July and October. 
For example, Fig.~\ref{fig:WaveHeight2019} shows the SWH when a typhoon approached Japan in 2019. 
The SWH on the Pacific Ocean side can reach approximately 10~m.

Fig.~\ref{fig:OceanCorr} shows the correlation coefficients of SWH at the 13 sites. 
It suggests that the ocean waves around KAGRA can be categorized into three areas: the Sea of Japan side, the Pacific side of the coast facing east (Pacific side east), and the Pacific side of the coast facing south (Pacific side south). As these three groups have little correlation, it can be understood that their behaviors are independent.

\section{Estimation of the microseismic motion from the ocean wave}\label{sec:main}
\begin{figure}[htbp] \centering
    \includegraphics[clip,width=14cm]{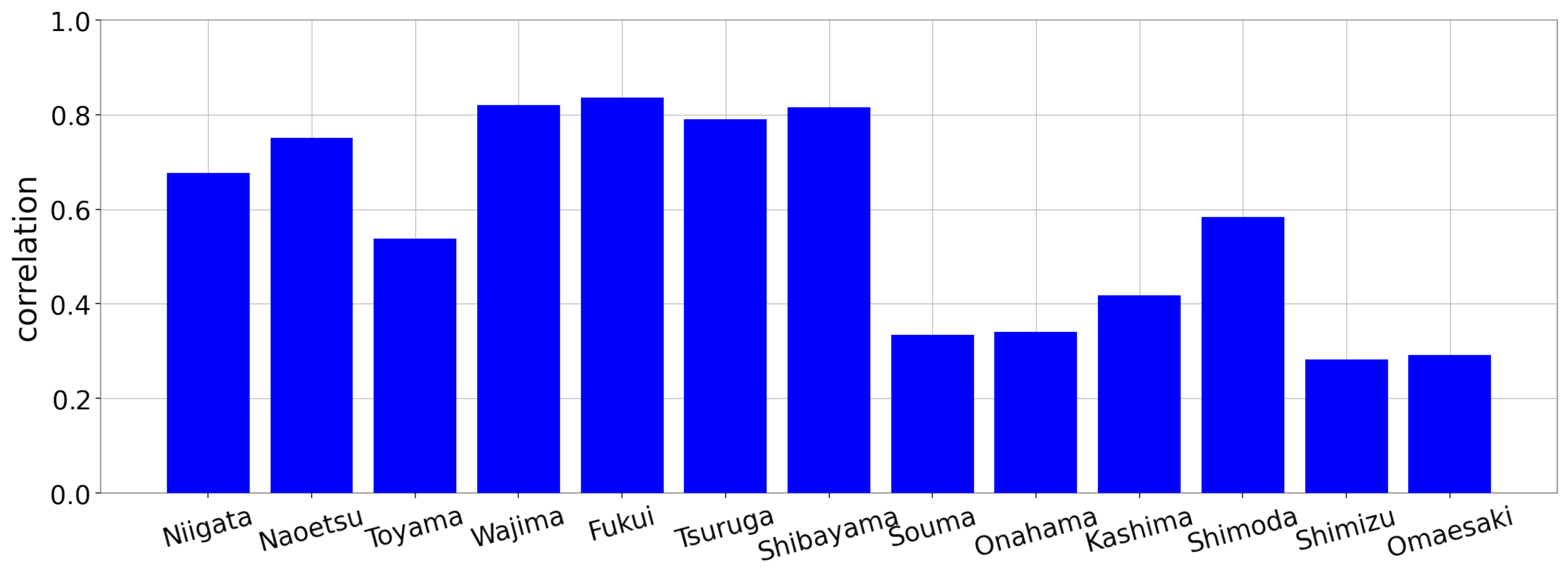}
    \caption{The Pearson correlation coefficient between the microseismic motion (BLRMS of the ground vibration velocity in 0.1--0.5 Hz) at the KAGRA site and the significant wave heights (SWH) in the NOWPHAS~\cite{NOWPHAS}, for the 13 coasts in Japan. 
The data is every 20 minutes in 2020.}
    \label{fig:SeisCorr}
\end{figure}

\begin{figure}[htbp] \centering
    \centering
    \includegraphics[clip,width=16cm]{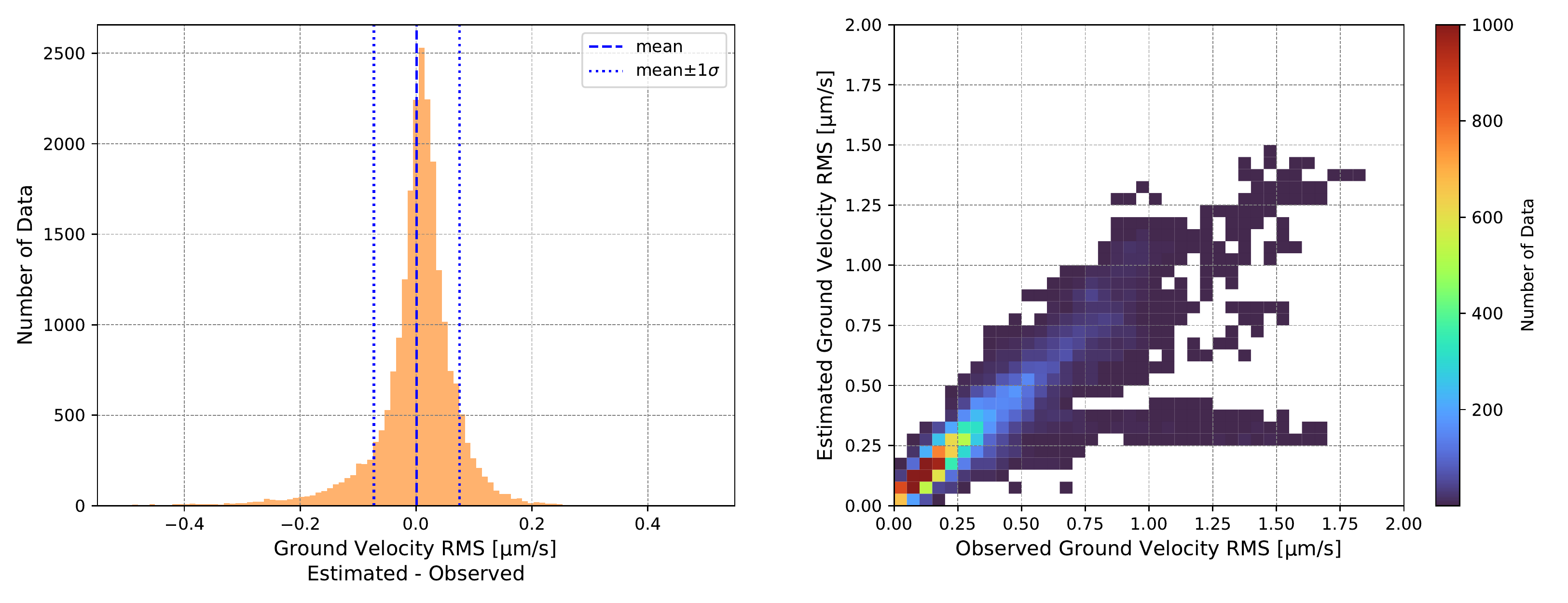}
    \caption{Left: Histogram of the difference between observed values and estimated values of the microseismic motion.
             Right: 2D histogram of the observed values and estimated values of the microseismic motion.}
    \label{fig:Sec4histgram}
\end{figure}

Data analysis utilizes seismometer data from KAGRA and ocean wave data. Both datasets were collected in 2020. Data during earthquake occurrence or approaching typhoons were excluded to ensure accuracy in the analysis.
To evaluate the relationship between microseismic motion at KAGRA and ocean waves, correlation analysis between seismometer signals and SWH at each coast was conducted. Fig.~\ref{fig:SeisCorr} shows correlation coefficients between the microseismic motion at the KAGRA site and the SWH data from NOWPHAS described in Sections~\ref{sec:seis} and ~\ref{sec:wave}. 
It indicates a strong positive correlation with the Sea of Japan and a weak positive correlation with the Pacific Ocean. 
This is expected as KAGRA is approximately 70~km from the Sea of Japan and 200~km from the Pacific Ocean at their closest points. We aim to establish a simple equation relating ground velocity BLRMS $v(t)$ and SWH $H_{1/3}(t)$.
For instance, Ferretti \textit{et al.} introduced the following equation:

\begin{equation}
    H_{1/3}(t) = \exp\Big[ a+b\ln v(t) \Big], 
\end{equation}
where constants $a = 9.48$ and $b = 0.66$ were derived through fitting~\cite{Prediction}. 
The objective of this section is to derive a similar equation between the microseismic motion at the KAGRA site and the SWH data in 2020 from NOWPHAS. Data during the typhoon period (Oct. 1 -- Oct. 15) were excluded, except for the final validation in Fig.~\ref{fig:Sec4histgram}.

\begin{figure}[htbp] \centering
    \includegraphics[clip,width=12cm]{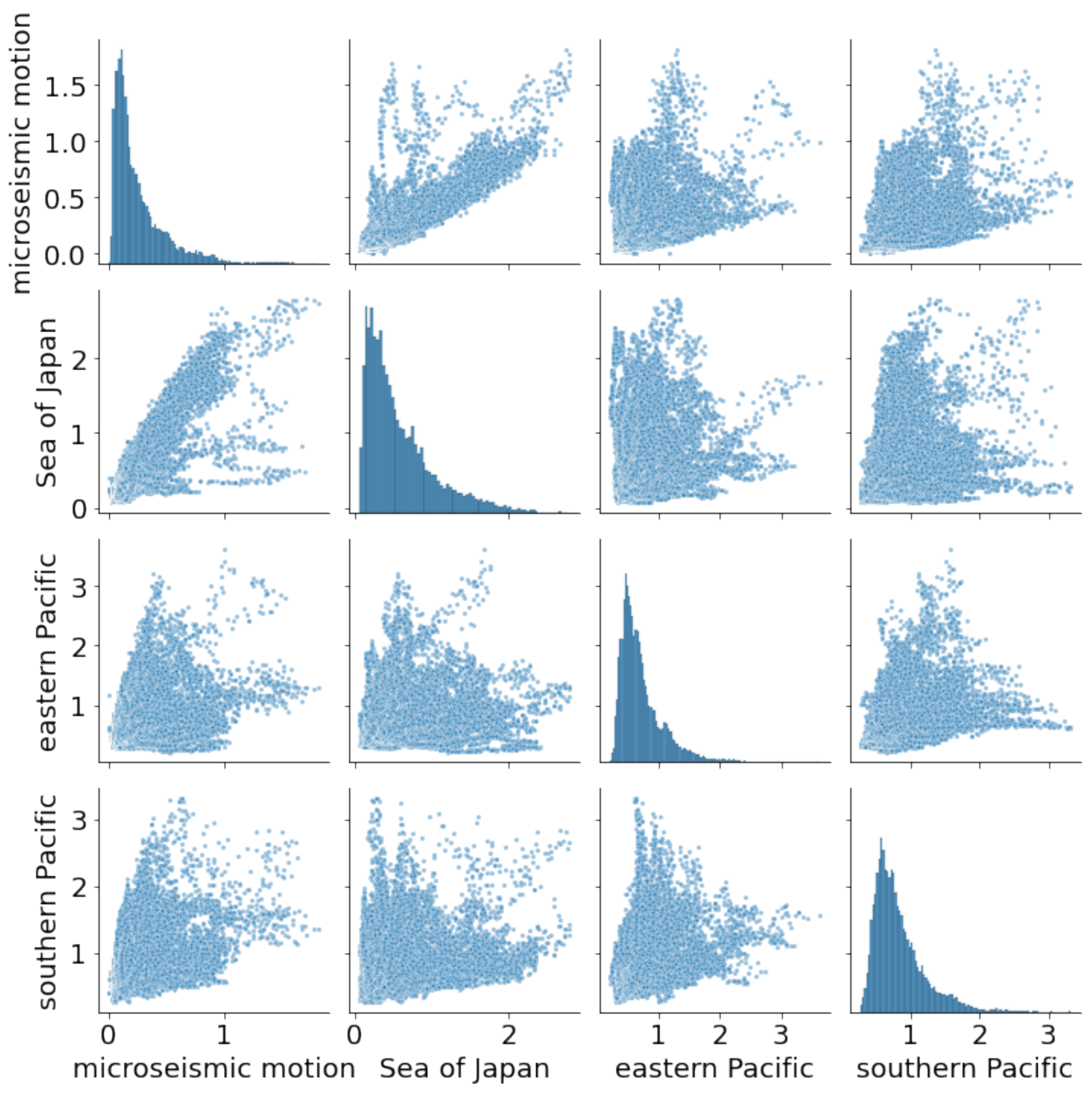}
    \caption{Scatter plot matrix of the observed microseismic motion at the KAGRA site and the representative wave levels for the Sea of Japan, the eastern Pacific side, and the southern Pacific side.}
    \label{fig:PCAScatter}
\end{figure}

\begin{figure}[htbp] \centering
    \includegraphics[clip,width=12cm]{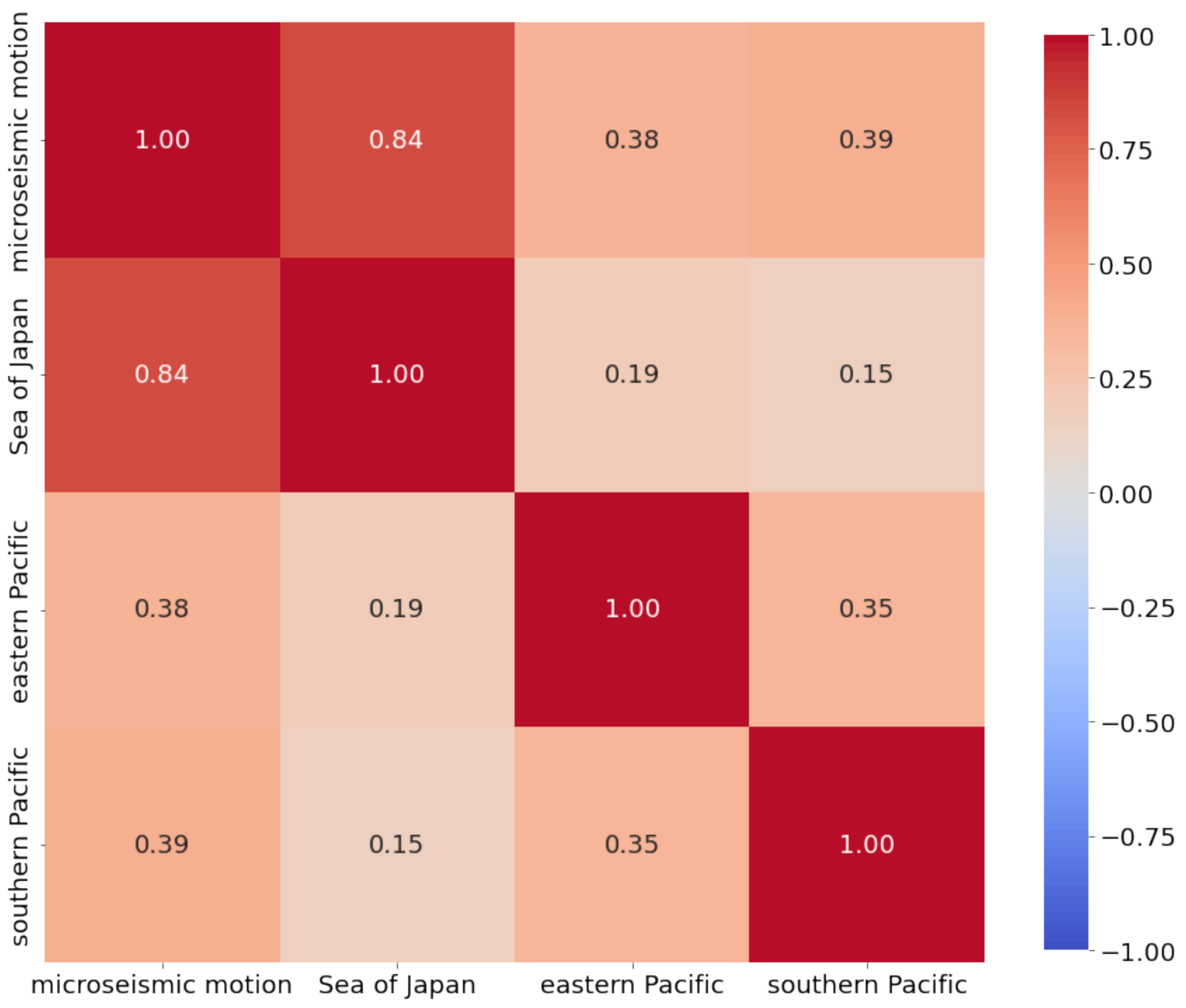}
    \caption{Correlation coefficient among the observed microseismic motion at the KAGRA site and the representative wave levels for the Sea of Japan, the eastern Pacific side, and the southern Pacific side.}
    \label{fig:PCACorr}
\end{figure}

Here, we estimate the microseismic motion at the KAGRA site using wave data from 12 locations excluding Toyama Bay in NOWPHAS. 
In this study, the ground velocity BLRMS $v$ and SWH $H_{1/3}$ are squared to calculate energy. 
SWH data from nearby locations have a strong positive correlation, as shown in Fig.~\ref{fig:OceanCorr}.
To address data degeneracy, principal component analysis (PCA) is applied to three areas of the ocean: the Sea of Japan side (Niigata, Naoetsu, Wajima, Fukui, Tsuruga, and Shibayama), the eastern Pacific side (Soma, Onahama, and Kashima), and the southern Pacific side (Shimoda, Shimizu, and Omaesaki). 
Each squared SWH $H_{1/3,j}^2(t)$ is standardized to have a mean of 0 and a standard deviation of 1. 
\begin{equation}
    x_j (t) = \frac{H_{1/3,j}^2(t) - \mu_j}{\sigma_j},
\end{equation}
where $x_j (t)$ is the standardized wave data, and $\mu_j$ and $\sigma_j$ are the mean and the standard deviation of $H_{1/3,j}^2(t)$ for the $j$--th NOWPHAS site.
PCA (using \texttt{scikit-learn 1.0.2}) is performed for three areas (Sea of Japan side, eastern Pacific side, and southern Pacific side) labeled by index $i$:
\begin{equation}
    PC_{i} (t) = \sum_{j} c_{ij} x_j (t),
\end{equation}
where $PC_{i} (t)$ is the first-principal component, and $c_{ij}$ is its eigenvector for the $i$-th area. 
The PCA parameter values are summarized in Table~\ref{tab:PCA}. 
\begin{table}[htbp] \centering
 \caption{Summary of the ocean wave data. $\mu$ and $\sigma$ represent the mean and standard deviation of the squared wave heights, respectively. $c_{ij}$ represents the PCA eigenvectors for three areas: the Sea of Japan side, the Eastern Pacific side, and the Southern Pacific side.}
  \label{tab:PCA}
  \begin{tabular}{l|ccccc} \hline
            &$\mu$ [m$^2$]  & $\sigma$ [m$^2$] &              &  $c_{ij}$  &   \\
            &        &                         & Sea of Japan &  Eastern Pacific &  Southern Pacific \\ \hline \hline
  Niigata   &   1.7  &   3.1   &     0.39     &  --           &  --    \\
  Naoetsu   &   1.8  &   3.1   &     0.42     &  --           &  --    \\
  Wajima    &   2.3  &   3.5   &     0.43     &  --           &  --    \\
  Fukui     &   2.3  &   3.9   &     0.42     &  --           &  --    \\
  Tsuruga   &   1.3  &   2.5   &     0.40     &  --           &  --    \\
  Shibayama &   2.2  &   3.3   &     0.39     &  --           &  --    \\ 
  Soma      &   1.2  &   1.8   &     --       &  0.58         &  --    \\
  Onahama   &   1.5  &   2.1   &     --       &  0.58         &  --    \\
  Kashima   &   2.4  &   3.3   &     --       &  0.57         &  --    \\
  Shimoda   &   1.0  &   1.0   &     --       &  --           &  0.58  \\
  Shimizu   &   0.26 &   0.32  &     --       &  --           &  0.59  \\
  Omaesaki  &   0.91 &   1.09  &     --       &  --           &  0.57  \\ \hline
  \end{tabular}
\end{table}
These principal components $PC_{i} (t)$ are difficult to compare to the microseismic motion because they are dimensionless and include negative values. 
Therefore, the weighted average of the squared SWH 
\begin{equation}
    \bar H_{i}^2(t) = \sum_{j} w_{ij} H_{1/3,j}^2(t), \quad w_{ij}=\frac{c_{ij}/\sigma_j}{\sum_{k}c_{ik}/\sigma_k},
\end{equation}
provides a better representation of the wave levels for each ocean area. 
Fig.~\ref{fig:PCAScatter} and Fig.~\ref{fig:PCACorr} are the correlation plots and the correlation coefficients for the observed microseismic motion $v_\mathrm{obs}^2(t)$ and the wave levels of the three ocean areas $\bar H_{i}^2(t)$. 
There was no strong correlation between the individual representative significant wave heights. 
They also showed that waves on the Sea of Japan side strongly influence the ground vibration values.

\begin{figure}[htbp] \centering
        \includegraphics[clip,width=8cm]{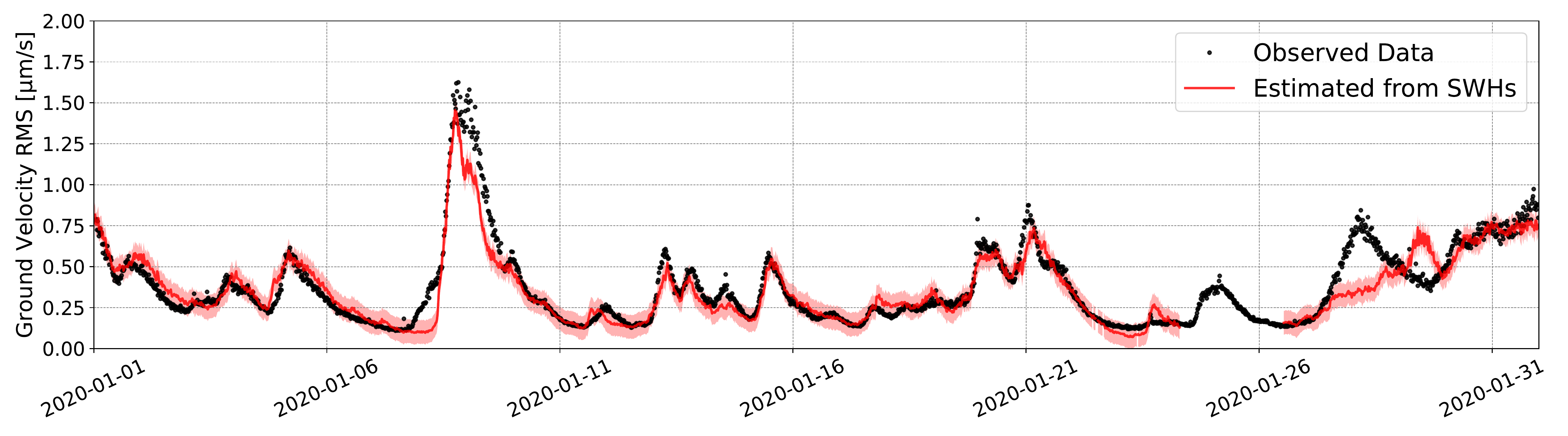}
        \includegraphics[clip,width=8cm]{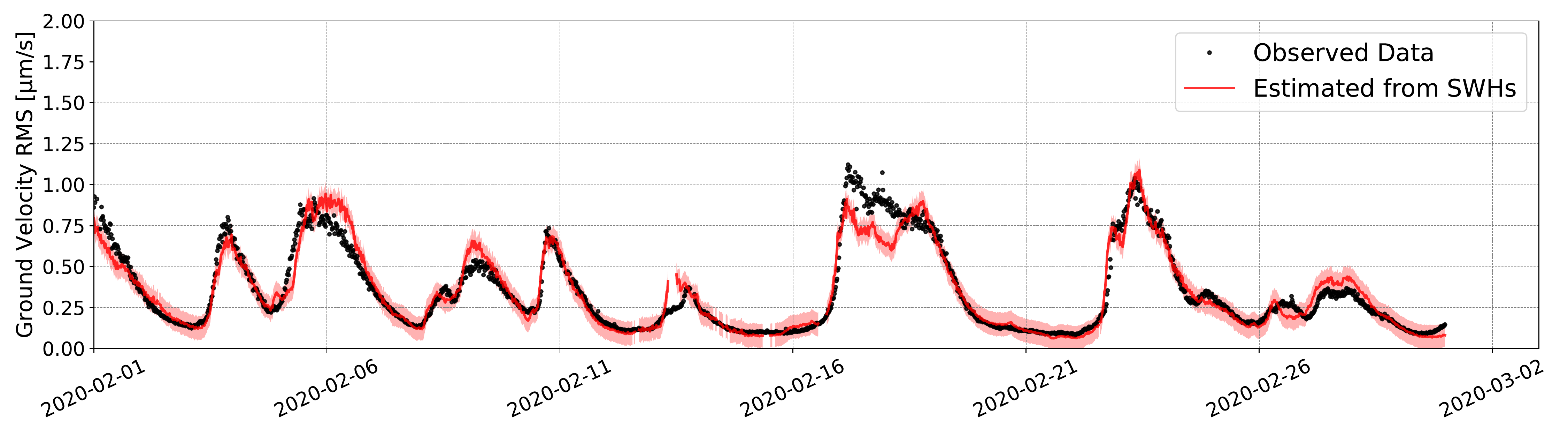}\\
        \includegraphics[clip,width=8cm]{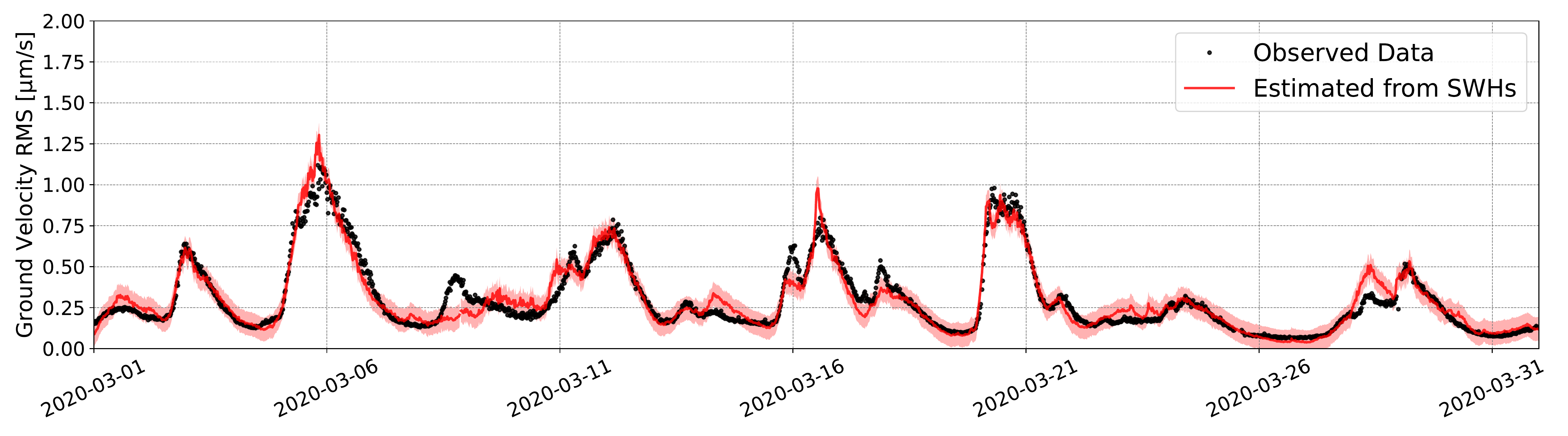}
        \includegraphics[clip,width=8cm]{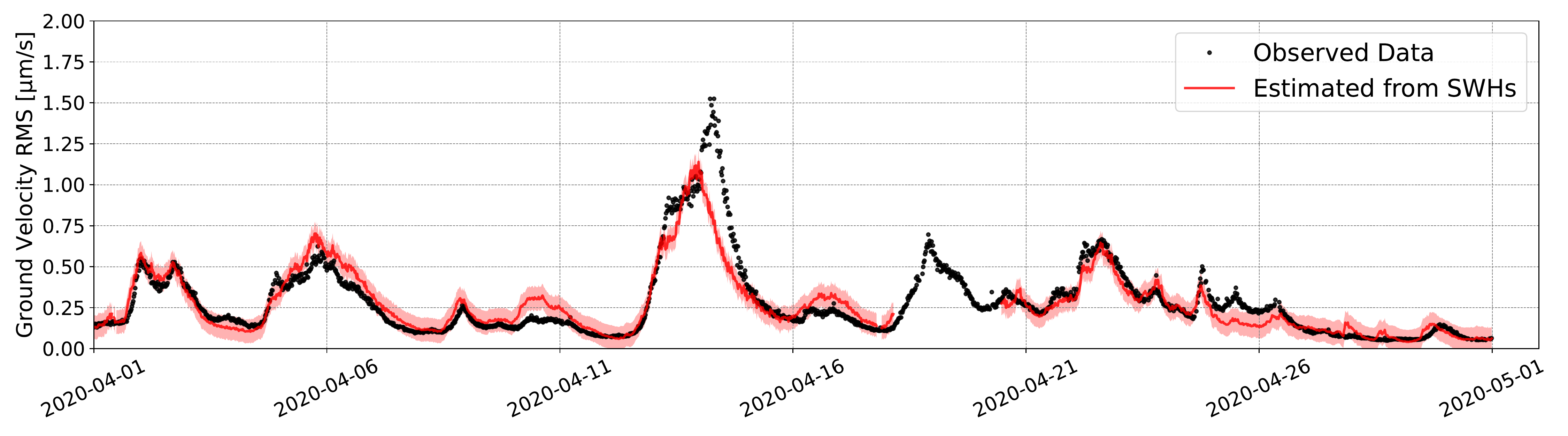}\\
        \includegraphics[clip,width=8cm]{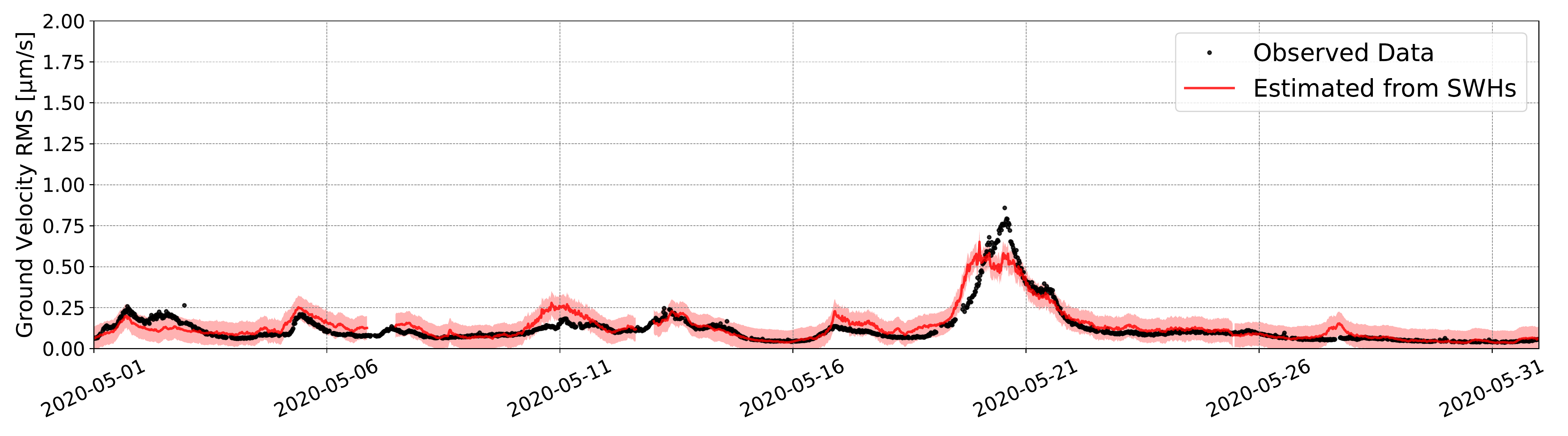}
        \includegraphics[clip,width=8cm]{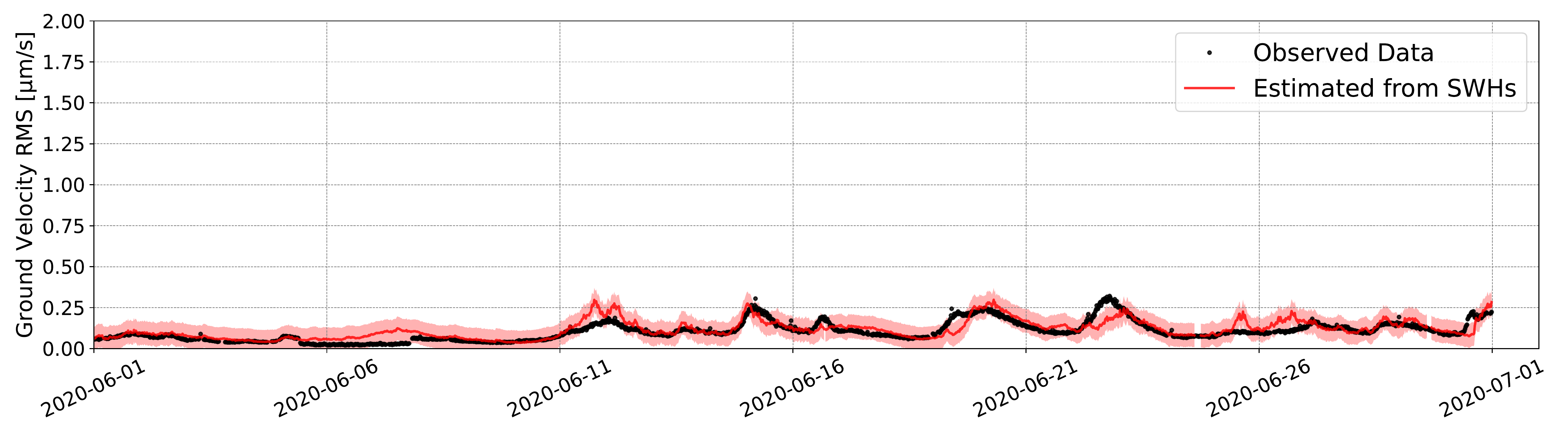}\\
        \includegraphics[clip,width=8cm]{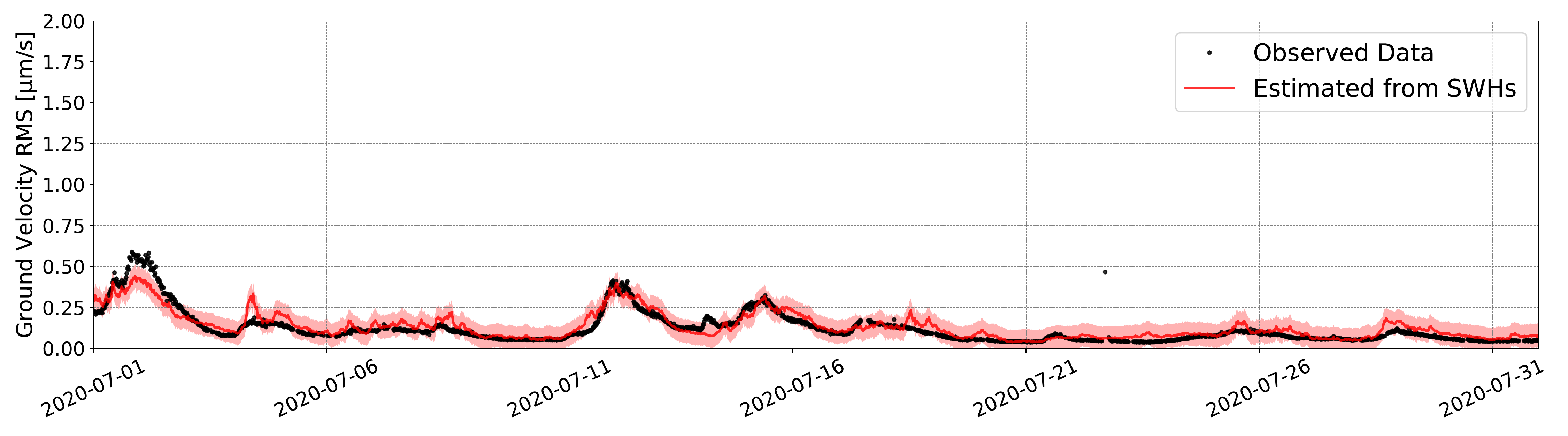}
        \includegraphics[clip,width=8cm]{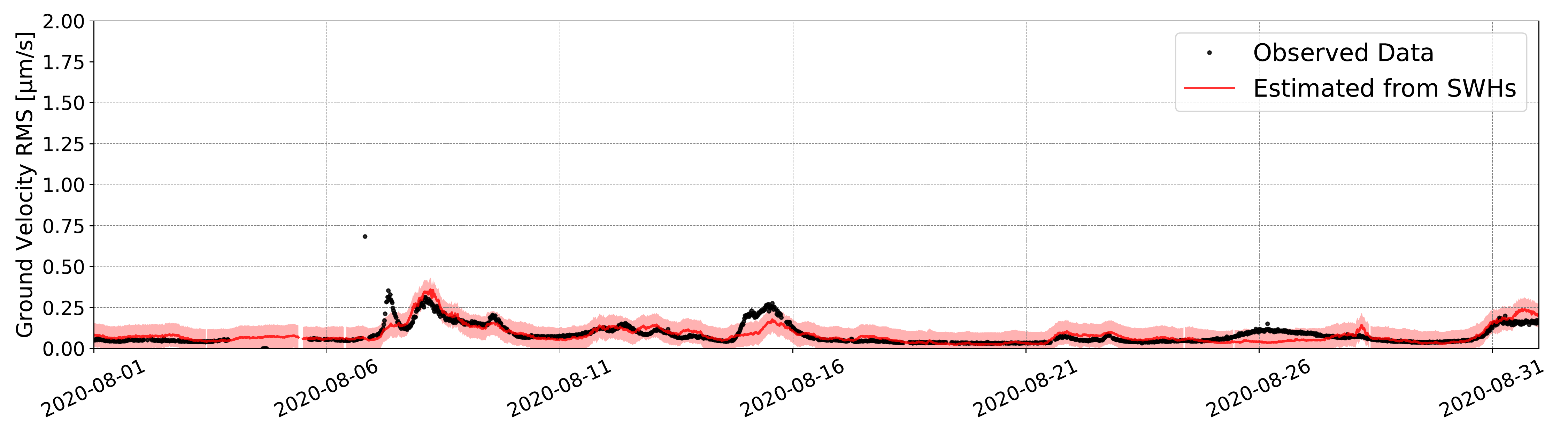}\\
        \includegraphics[clip,width=8cm]{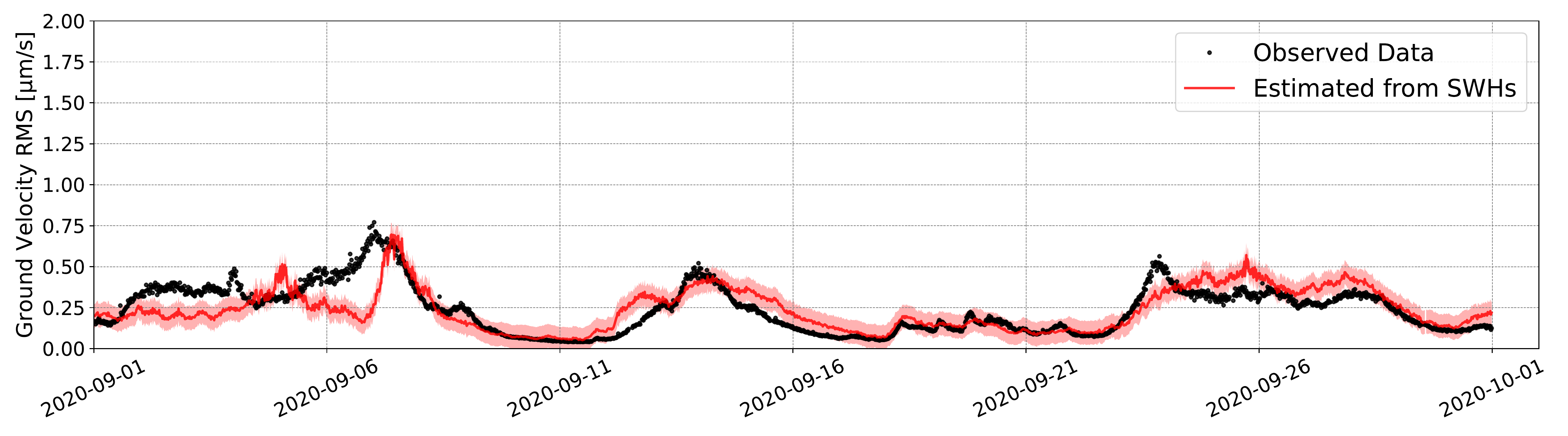}
        \includegraphics[clip,width=8cm]{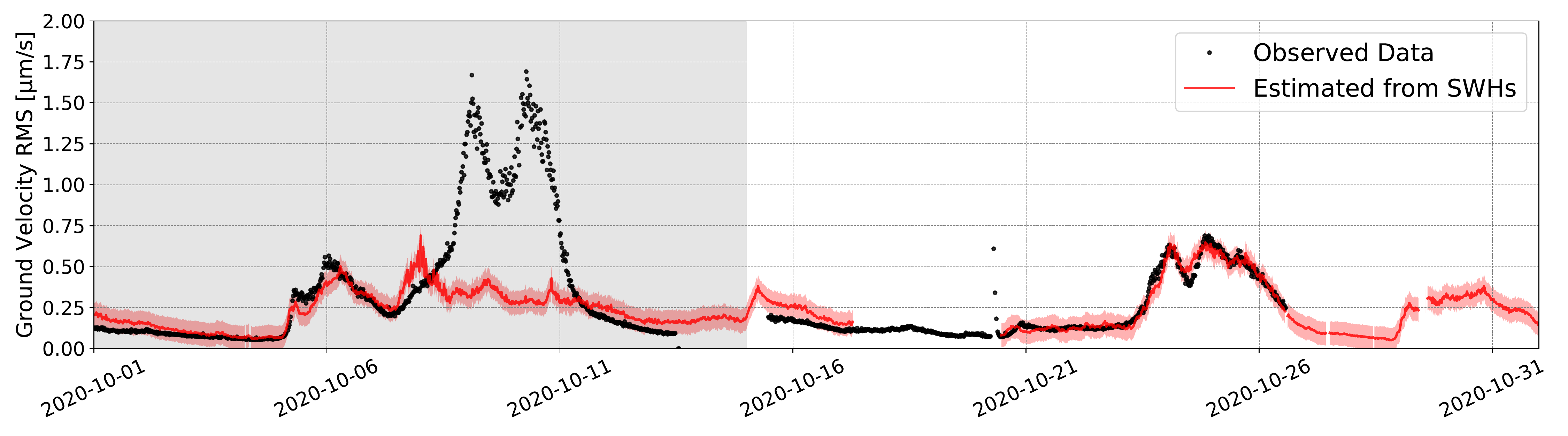}\\
        \includegraphics[clip,width=8cm]{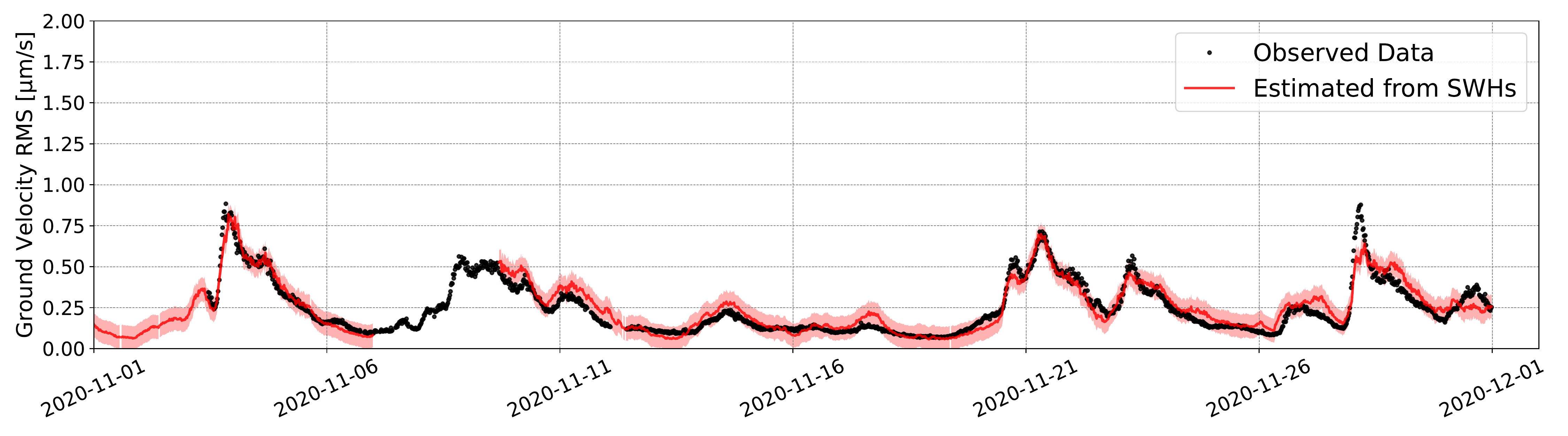}
        \includegraphics[clip,width=8cm]{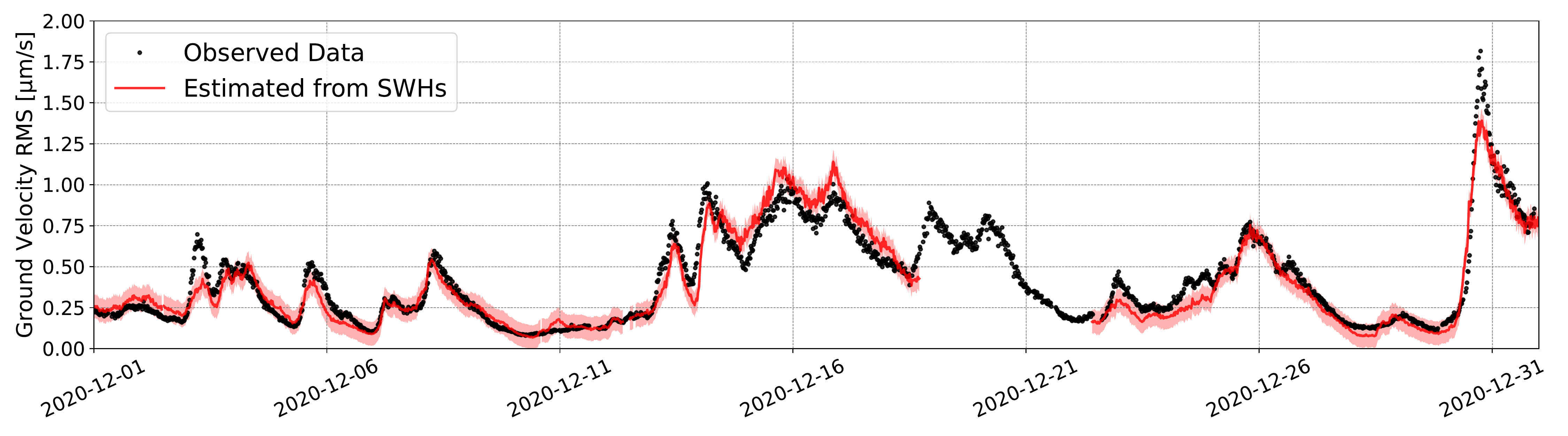}
    \caption{Comparison of the microseismic motion at the KAGRA site, between the observed data (black) and the estimation from the ocean waves (red, with the $1\sigma$ error band).
    The typhoon period (Oct. 1 -- Oct. 15, gray hatched) is not used for the estimation.}
    \label{fig:Sec4PredTimeseries}
\end{figure}

The estimation for the microseismic motion at the KAGRA site is written as follows:
\begin{equation}
    v_\mathrm{est}^2(t) = \sum_{i} \alpha_i^2 \bar H_{i}^{2\beta_i} (t), \label{eq:fit}
\end{equation}
where $\alpha_i$ and $\beta_i$ are constants and are derived by fitting the data using the non-linear least squares method (via \texttt{scipy.optimize.curve\_fit}). 
The results are shown in Fig.~\ref{fig:Sec4PredTimeseries}, where the red lines represent the estimated microseismic motion $v^2_\mathrm{est}(t)$ from the NOWPHAS data, and the black markers indicate the observed microseismic motion $v^2(t)$ for one year in 2020. The values of the fitting parameters are shown in Table~\ref{tab:fit_result}. 
There is almost good agreement between the estimation and the observed data, except for the typhoon data not used in the PCA and the fitting. 
The values of the index $\beta_i$ are close to $b^{-1} \sim 1.5$ of Ferretti {\it et al.}~\cite{Prediction}.

\begin{table}[htbp] \centering
    \caption{The results of the fitting parameters in Eq.~(\ref{eq:fit}). }
    \label{tab:fit_result}   
    \begin{tabular}{c|cc}    \hline
                       & $\alpha_i\ \mathrm{[\mu m/s/m^{\beta_i}]}$  & $\beta_i$ \\ \hline\hline
        Sea of Japan   & $0.358  \pm 0.001$                              & $1.314 \pm 0.005$ \\
        Eastern Pacific   & $0.104  \pm 0.002$                              & $1.644 \pm 0.024$ \\
        Southern Pacific  & $0.092  \pm 0.002$                              & $1.687 \pm 0.024$ \\ \hline
    \end{tabular}
\end{table}

To discuss the accuracy of this estimation, Fig.~\ref{fig:Sec4histgram} (left) shows a histogram of the difference between the estimated and observed microseismic motion. 
It is almost a Gaussian distribution with a mean and standard deviation of $1.2\times10^{-3}\mathrm{\mu m/s}$ and $7.4\times10^{-2}\mathrm{\mu m/s}$, respectively. 
This standard deviation was used for the error band in Fig.~\ref{fig:Sec4PredTimeseries}.
Fig.~\ref{fig:Sec4histgram} (right) shows a 2D histogram of the data points between the estimated and observed microseismic motions, confirming good linearity for most parts.
This accuracy is achieved owing to the index parameter $\beta_i$; if not used (corresponding to $\beta_i=1$), the estimation becomes systematically smaller when the ocean waves are violent.
The cluster distributes around $v_\mathrm{est}\sim 0.3\ \mathrm{\mu m/s}$ and $v_\mathrm{obs} > 0.7 \ \mathrm{\mu m/s}$ corresponding to the period when the typhoon was approaching Japan (during October 9--11). 
This is because an offshore typhoon directly shakes the seabed just below it, and microseismic waves propagate to Japan even though the coasts are still not rough~\cite{typhoon}.

\section{Conclusion and prospects}\label{sec: prospects}
In this study, we investigated the properties of the microseismic motion at the KAGRA site and the significant wave heights on the coasts of Japan measured and released by NOWPHAS for 2020. 
The degeneracy of the wave data was resolved using a principal component analysis, and the first principal components of the three ocean regions were extracted. 
Microseismic motion observed at the KAGRA site was estimated by the combination of these components of the ocean waves within the standard deviation of $7\times10^{-2}\mathrm{\mu m/s}$.

In this paper, only the first principal components for each sea area are used, and other principal components (which contain the local information) are not considered. Models that incorporate all information may provide better estimation accuracy.
Other kinds of data, such as significant wave periods and wind direction, are also important. 
Machine learning is a possible way to improve this analysis. 
To include typhoon days, the development of special treatments is necessary, \textit{e.g.}, using the position and magnitude of the typhoon.

\medskip

\begin{figure}[htbp] \centering
    \includegraphics[clip,width=15cm]{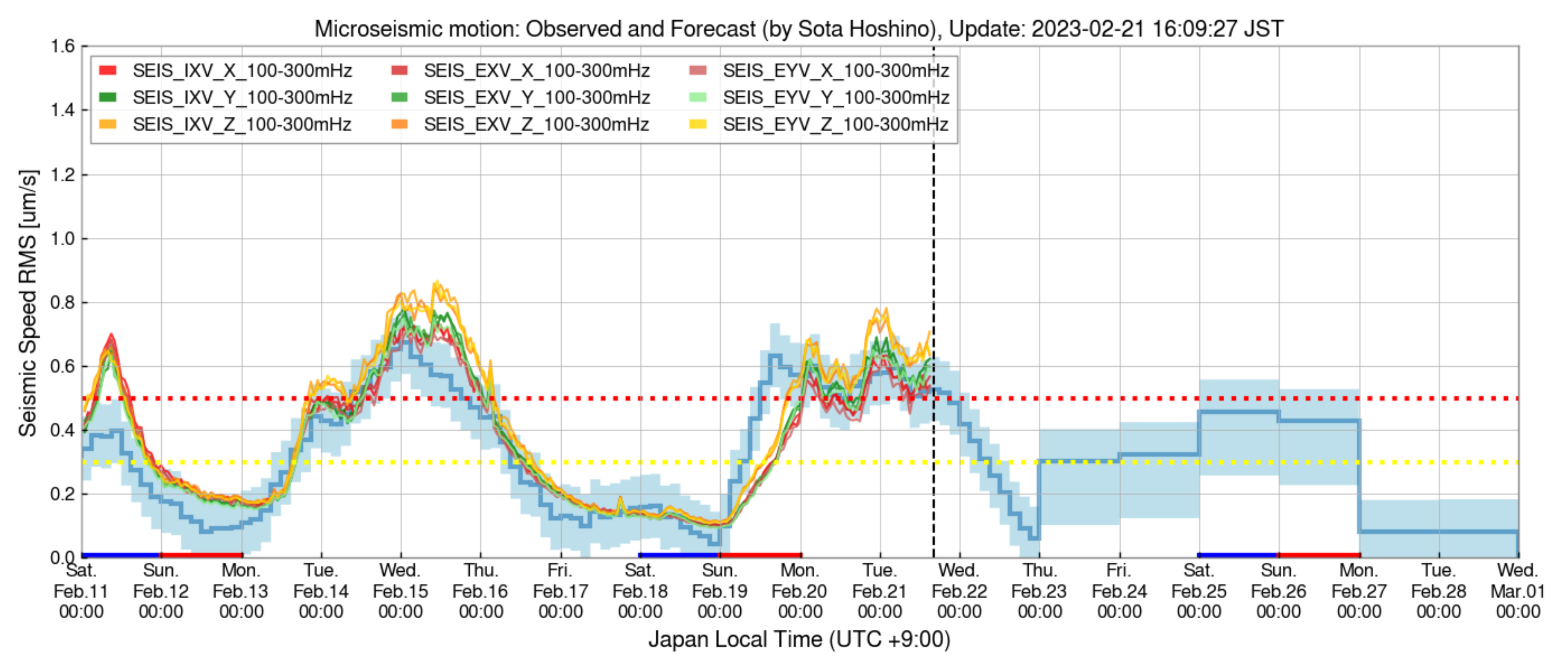}
    \caption{An example of the microseismic forecast graph. The black line shows the current date, in this case, 2023-05-18 11:09 (JST). The right and left sides of it are the future and past, respectively. The horizontal dotted lines (red and yellow) correspond to the benchmark microseismic level discussed in Sec.~\ref{sec:seis}, red is $0.5~\mathrm{\mu m/s}$, and yellow is $0.3~\mathrm{\mu m/s}$. The historical data is overlaid with BLRMS of the actual measured triaxial ground vibrations from the three seismometers located at the KAGRA site for comparison.}
    \label{fig:MicroseisForecast}
\end{figure}

A useful application of this study is \textit{the microseismic forecast}. 
Future microseismic motion can be forecasted by inputting wave information from commercial weather forecasts into our equation. 
Fig.~\ref{fig:MicroseisForecast} shows an example of a microseismic forecast using a 1-week weather forecast on the website \texttt{Otenki.com} provided by Bellsystem24 Inc~\cite{Otenki.com}. 
The blue line represents the forecast with the error band of microseismic motion at the KAGRA site.
The vertical black dotted line represents the current time, and the right and left-hand sides represent the future and past, respectively. 
The horizontal dotted lines (red and yellow) correspond to the benchmark microseismic levels discussed in Section~\ref{sec:seis}. 
The record of the historical forecast value is overlaid with BLRMS of the actual measured values, allowing users to assess the accuracy of the recent forecast.
This plot is automatically updated and shown on a website (internal page) and contributes to commissioning works at the KAGRA observatory, especially for its scheduling.

\section*{Acknowledgement}

This research has made use of data, software, and web tools obtained or developed by the KAGRA Collaboration, especially the administrators of the digital system, and managers of the KAGRA site. 
The NOWPHAS data is provided by the Port and Harbor Bureau, Ministry of Land, Infrastructure, Transport and Tourism, Japan, and the weather forecast data is provided by Bellsystem24 Inc.

The KAGRA project is supported by 
MEXT, 
JSPS Leading-edge Research Infrastructure Program, 
JSPS Grant-in-Aid for Specially Promoted Research 26000005, 
JSPS Grant-in-Aid for Scientific Research on Innovative Areas 2905: JP17H06358, JP17H06361 and JP17H06364, 
JSPS Core-to-Core Program A. Advanced Research Networks, JSPS Grantin-Aid for Scientific Research (S) 17H06133 and 20H05639, 
JSPS Grant-in-Aid for Transformative Research Areas (A) 20A203: JP20H05854, 
the joint research program of the Institute for Cosmic Ray Research, University of Tokyo, 
National Research Foundation (NRF), 
Computing Infrastructure Project of Global Science experimental Data hub Center (GSDC) at KISTI, 
Korea Astronomy and Space Science Institute (KASI), 
and Ministry of Science and ICT (MSIT) in Korea, 
Academia Sinica (AS), 
AS Grid Center (ASGC) 
and the National Science and Technology Council (NSTC) in Taiwan under grants including the Rising Star Program and Science Vanguard Research Program, 
Advanced Technology Center (ATC) of NAOJ, and Mechanical Engineering Center of KEK. 
Especially this work was founded by
JSPS Grant-in-Aid for JSPS Fellows 19J01299 
and the Joint Research Program of the Institute for Cosmic Ray Research (ICRR) University of Tokyo 2020-G07, 2020-G12, 2020-G21, 2021-G07, 2021-G09, 2021-G10, 2022-G07, 2022-G12, 2022-G21.

We would like to thank Editage (\url{www.editage.com}) for English language editing.


\appendix
\section{One year data of the SWH and the microseismic motion}\label{sec:OneYear}
Appendix A contains SWH time series data for the full year 2020 that were not shown in Fig.~\ref{fig:Sec3TimeSeries}. In addition, this figure overlays the BLRMS of the KAGRA seismometer data.

\begin{figure}[p]  \centering
        \includegraphics[clip,width=14cm]{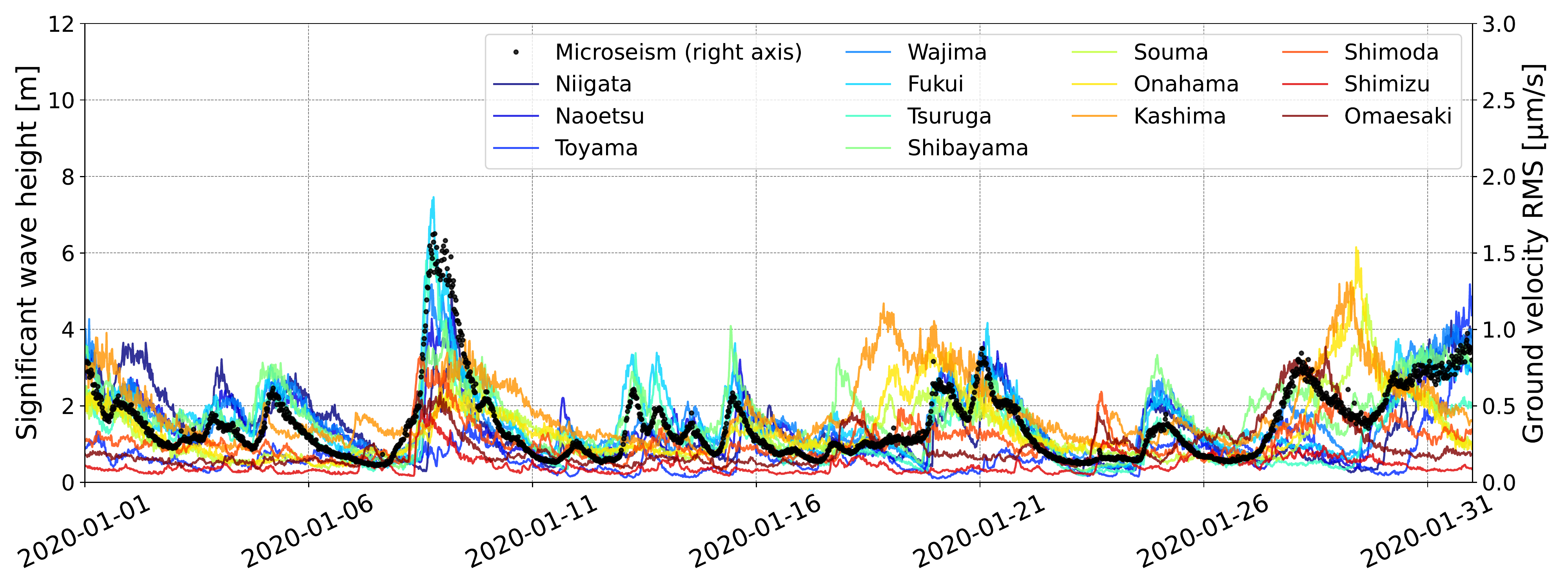}
        \includegraphics[clip,width=14cm]{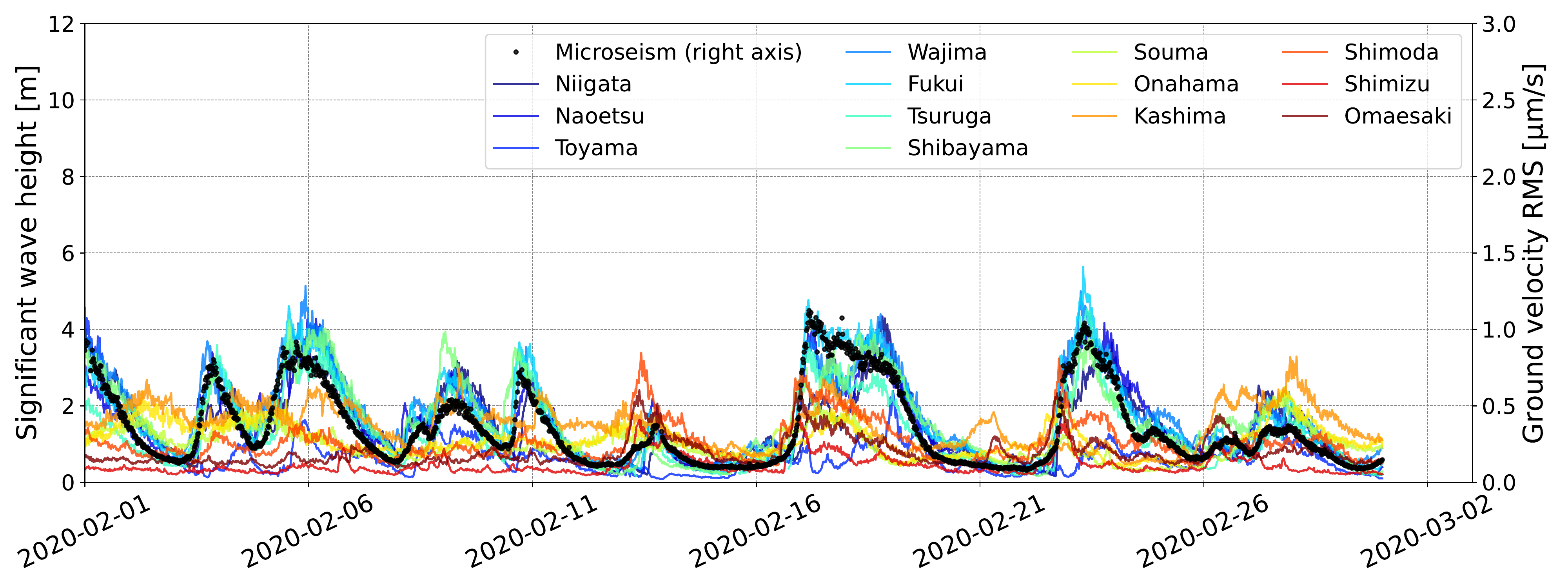}
        \includegraphics[clip,width=14cm]{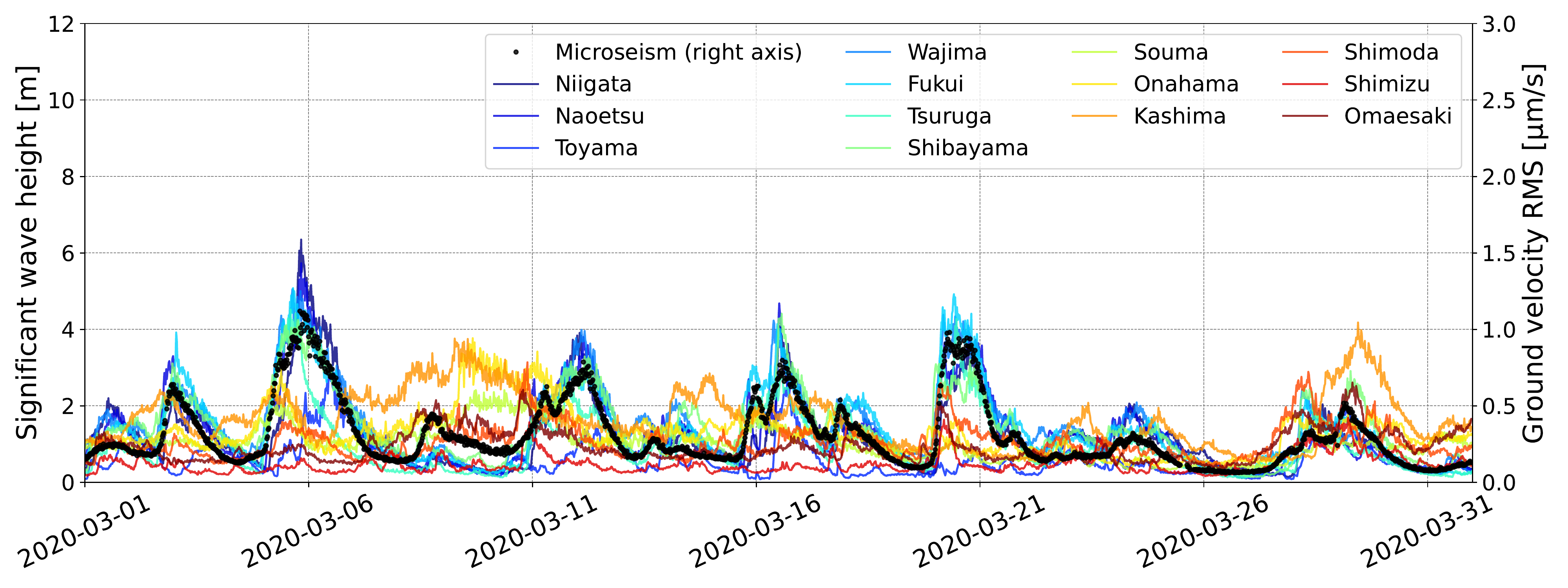}
        \includegraphics[clip,width=14cm]{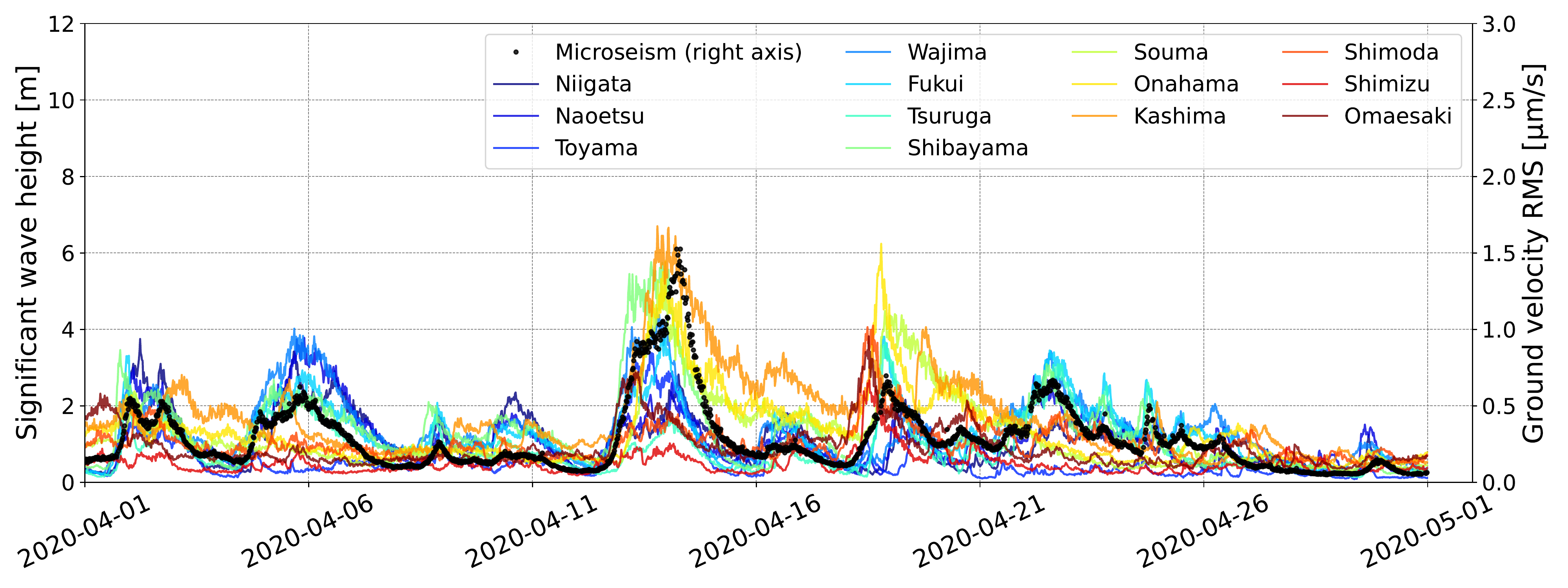}
\end{figure}

\begin{figure}[p]  \centering
        \includegraphics[clip,width=14cm]{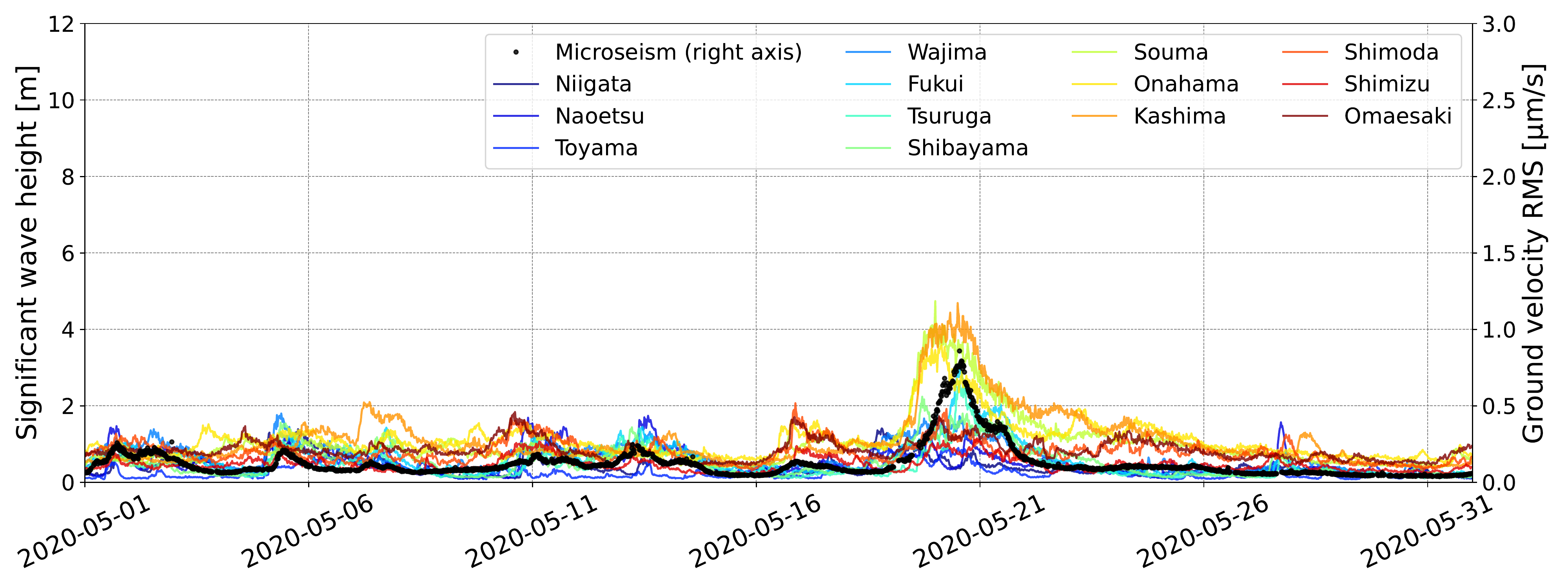}        \includegraphics[clip,width=14cm]{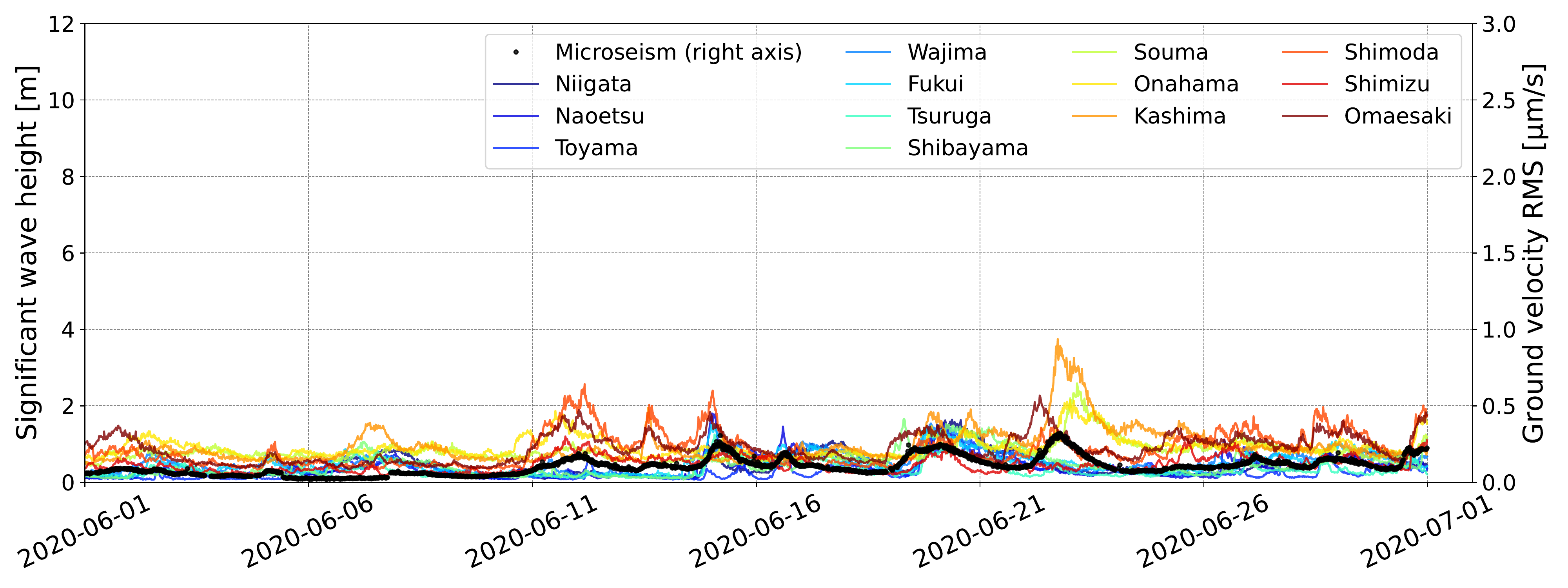}
        \includegraphics[clip,width=14cm]{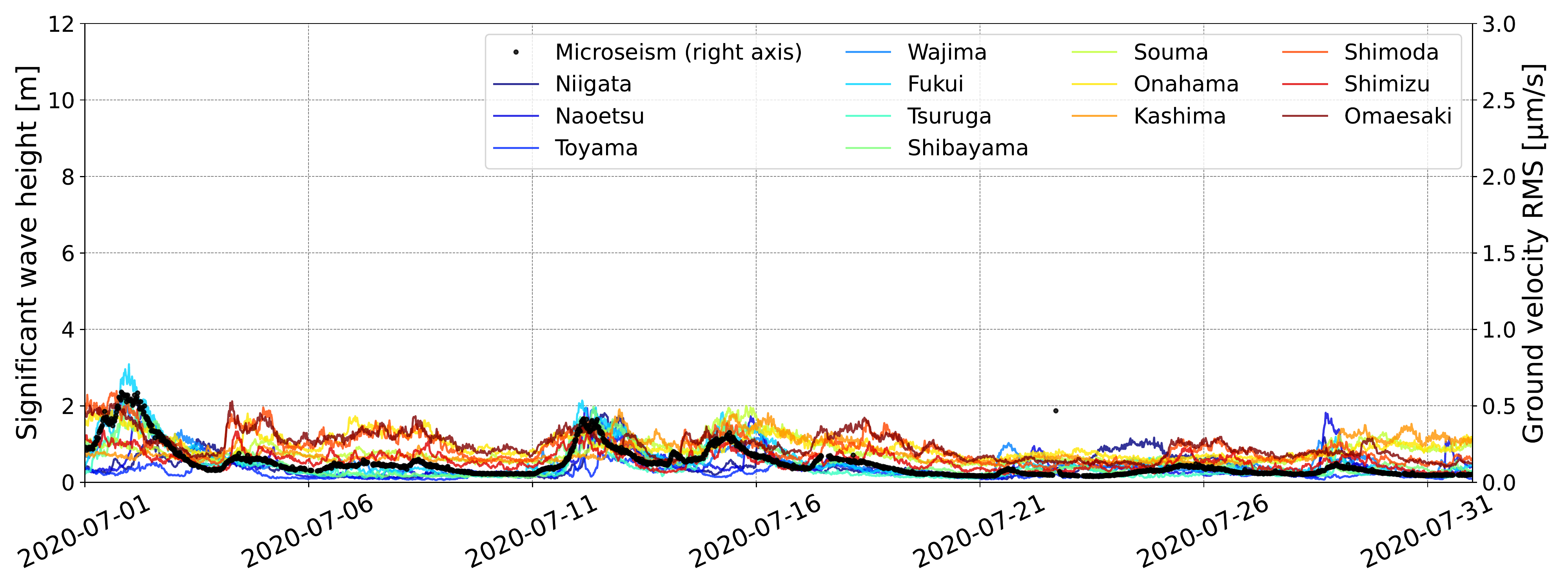}
        \includegraphics[clip,width=14cm]{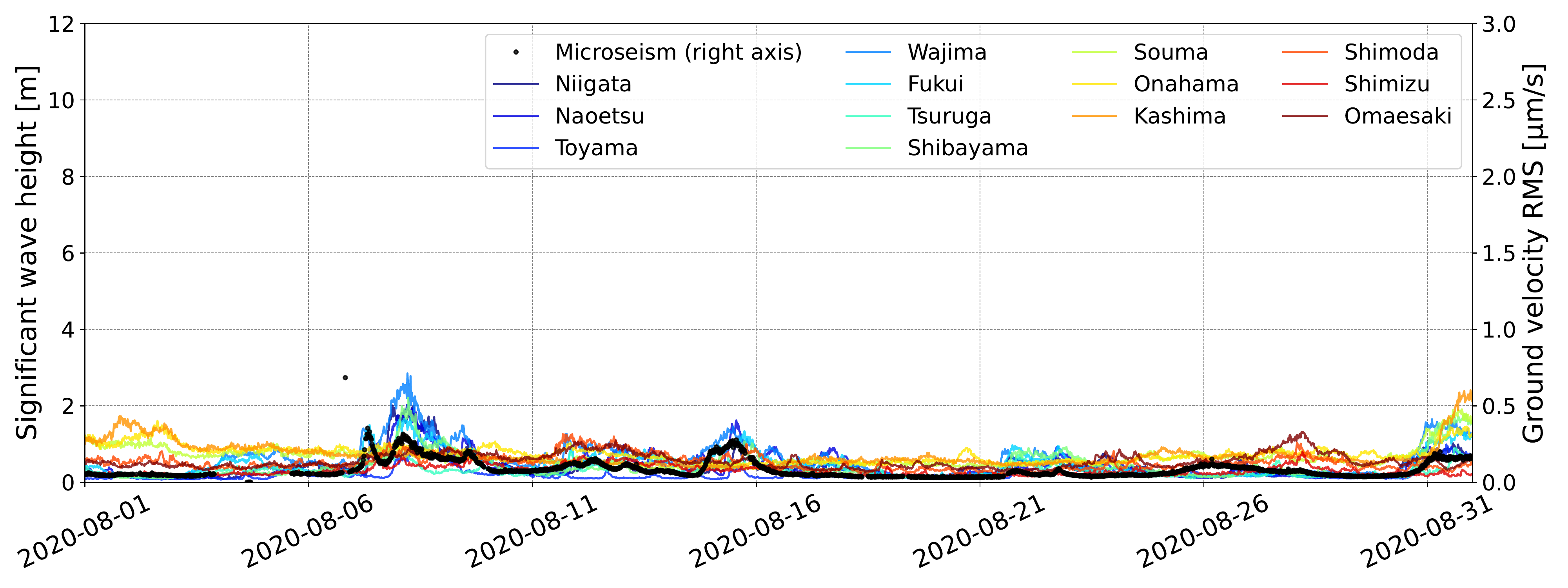}
\end{figure}

\begin{figure}[p]  \centering
        \includegraphics[clip,width=14cm]{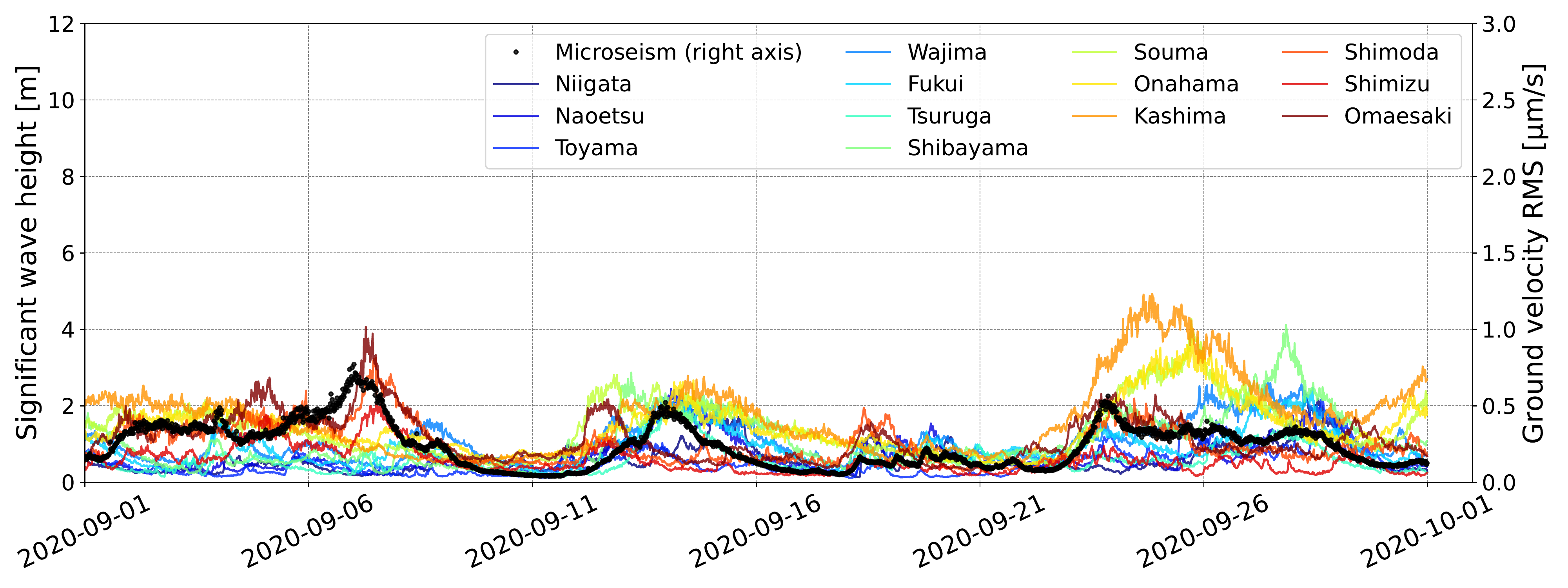}
        \includegraphics[clip,width=14cm]{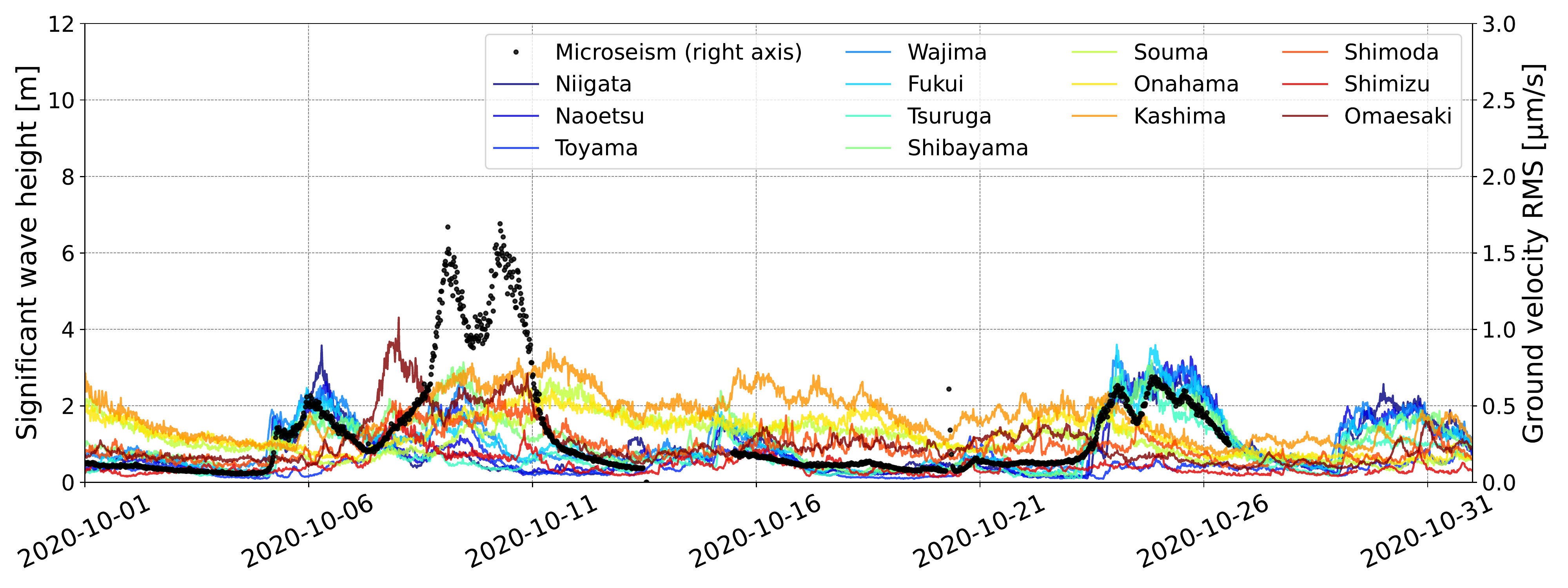}
        \includegraphics[clip,width=14cm]{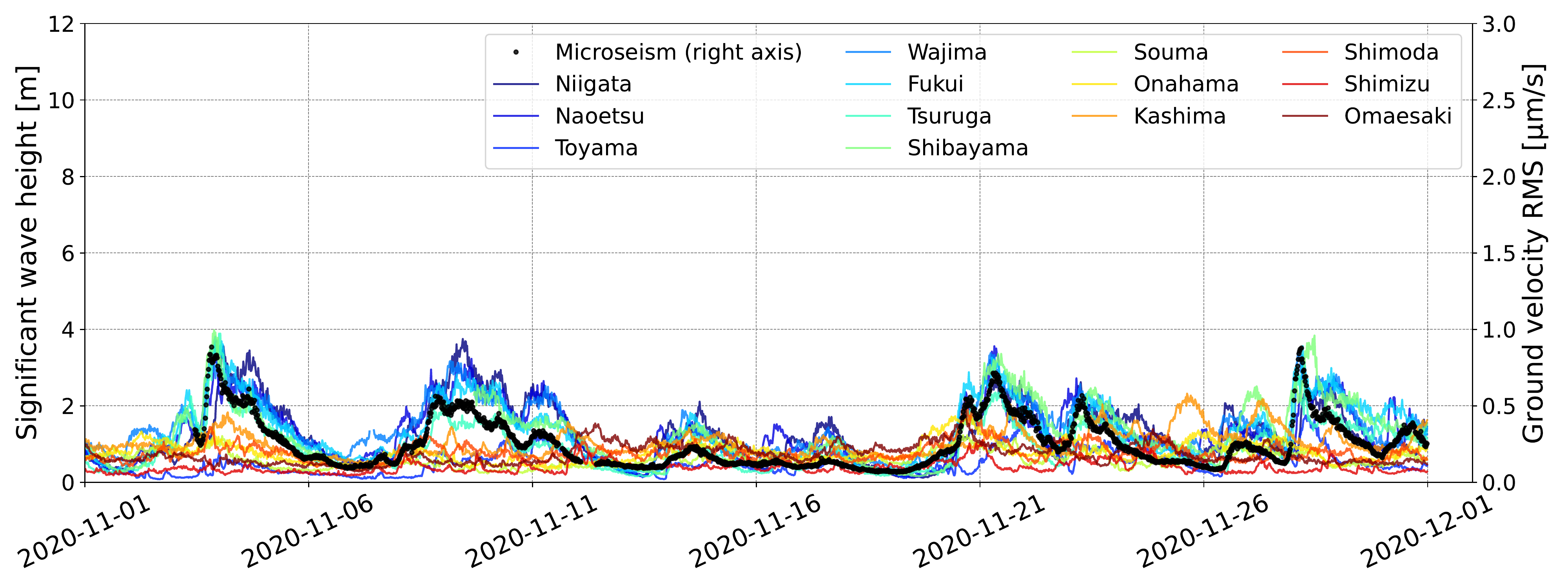}
        \includegraphics[clip,width=14cm]{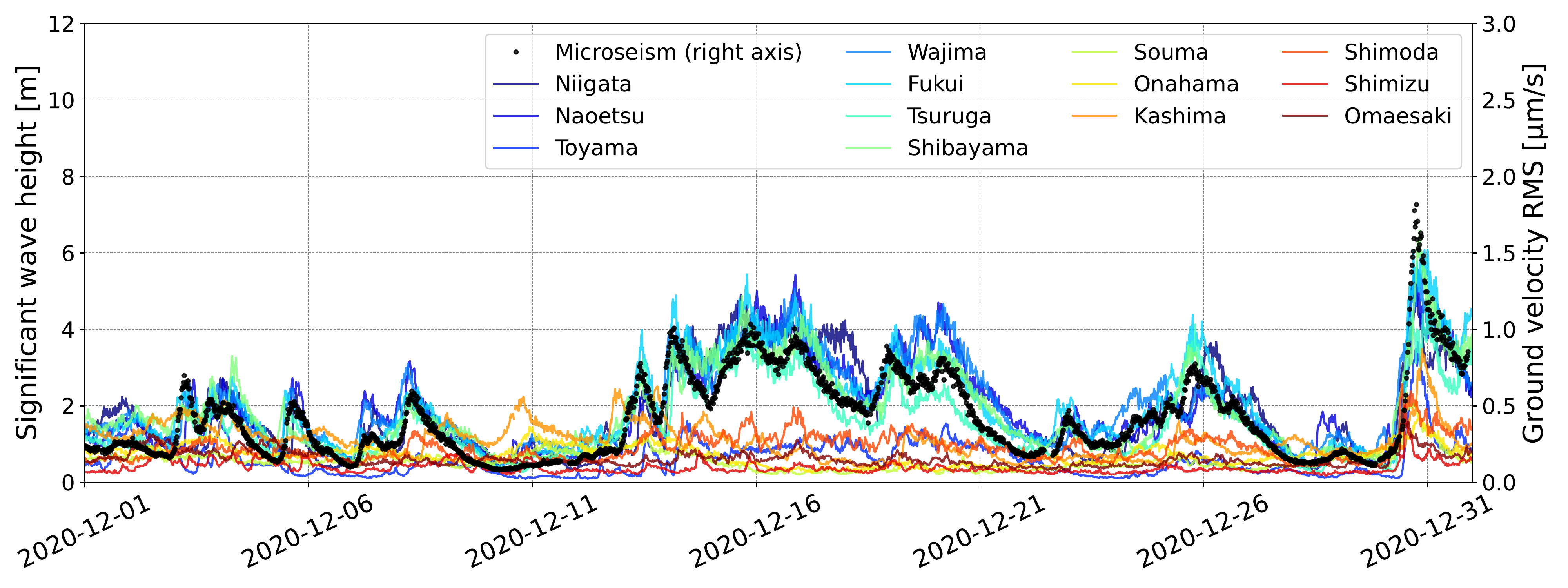}
    \caption{ The time series data of the SWH in NOWPHAS and the microseismic motion at the KAGRA site, partially shown in Fig.~\ref{fig:Sec3TimeSeries}, are shown here through the full year 2020.}
    \label{fig:Sec4SeisTimeseries}
\end{figure}

\end{document}